\documentclass[%
twocolumn,
 floatfix,
 superscriptaddress,
nofootinbib,
 amsmath,amssymb,
 aps,
]{revtex4-2}

\usepackage{booktabs}
\usepackage[x11names]{xcolor}
\usepackage{graphicx}
\usepackage{dcolumn}
\usepackage{bm}
\usepackage{hyperref}
\usepackage{cleveref}
\usepackage[mathlines]{lineno}
\usepackage{todonotes}
\usepackage{subcaption}
\usepackage[normalem]{ulem}
\usepackage{ragged2e}
\usepackage[x11names]{xcolor}
\usepackage{custom}

\newcommand{\coef}{\alpha}

\allowdisplaybreaks
\begin{document}

\preprint{APS/123-QED}

\title{Lattice-based equation of state with a critical point from constant entropy contours and its comparison to effective QCD approaches
}

\author{Hitansh Shah}
\affiliation{
 Department of Physics, University of Houston, Houston, TX 77204, USA
}

\author{Mauricio Hippert}
\affiliation{Instituto de Física, Universidade do Estado do Rio de Janeiro, Rua São Francisco Xavier, 524, Rio de Janeiro, RJ, 20550-013, Brazil}

\author{Jorge Noronha}
\affiliation{Illinois Center for Advanced Studies of the Universe \& Department of Physics,
University of Illinois at Urbana-Champaign, Urbana, IL 61801-3003, USA}

\author{Claudia Ratti}
\affiliation{
 Department of Physics, University of Houston, Houston, TX 77204, USA
}

\author{Volodymyr Vovchenko}
\affiliation{
 Department of Physics, University of Houston, Houston, TX 77204, USA
}

\date{\today}%

\begin{abstract}

In this work, we systematically assess the performance of a new method from [H. Shah et al., \href{https://doi.org/10.1103/cbwj-4jfl}{Phys. Rev. C 113, L012201 (2026)}] for locating the QCD critical point using constant-entropy contours by testing it against various effective QCD approaches.
We demonstrate that, while the method yields spurious critical points in purely hadronic models (HRG) due to non-parabolic contour behavior at low temperatures ($T \lesssim 120$ MeV), it accurately reproduces the CP location in frameworks that feature a genuine phase transition and benchmarked against lattice QCD, such as Holographic Einstein-Maxwell-Dilaton, and Functional QCD approaches.
Building on our previous determination of constant entropy contours using lattice data, we extend that analysis to construct a complete Lattice-based Equation of State (EoS) at finite density, which features a critical point at $(T, \mu_B) \approx (114, 602)$ MeV. 
By integrating the extrapolated entropy density with respect to temperature, we reconstruct the pressure, baryon density, susceptibility, and speed of sound in the critical region, and analyze the focusing behavior of isentropic trajectories in the vicinity of the critical point.

\end{abstract}

\maketitle

%
%
%
\section{Introduction} \label{sec:intro}
The existence of the conjectured QCD critical point (CP) has remained a mystery for decades now \cite{Bzdak:2019pkr,Du:2024wjm}. It is well known from first principles lattice QCD (LQCD) calculations that there exists a crossover from confined hadrons to a deconfined quark gluon plasma (QGP)  at vanishing chemical potential \cite{Aoki:2006br}.  The analytic crossover at zero baryon chemical potential occurs at a temperature of 155 -- 159 MeV \cite{Borsanyi:2020fev, HotQCD:2018pds}. 
Whether this smooth crossover turns into a first order phase transition at finite chemical potential is unknown, as direct lattice QCD simulations at nonzero baryon chemical potential are hindered due to the fermion sign problem.

Experimentally, the QCD critical point search has been the primary focus for the Relativistic Heavy Ion Collider (RHIC) for many years. The search will continue at the future Facility For Antiproton and Ion Research (FAIR) in Germany. Although no conclusive statement has been obtained on the location of the CP, the data hints at interesting behavior for energies $\sqrt{s_{NN}} \leq 20$ GeV \cite{STAR:2025zdq,STAR:2020tga,STAR:2021iop}, especially the new BESII data for proton cumulants and transverse momentum fluctuations. However, there are still sizable errors at the moment that do not allow us to confirm the location (or even existence) of the QCD critical point.

On the theoretical front, the location of the QCD critical point (CP) cannot be directly determined from lattice QCD simulations because of the fermion sign problem. As a result, several complementary approaches have been developed to extract the equation of state of strongly interacting matter at finite baryon chemical potential. These include analytic continuation from imaginary $\mu_B$, searches for Lee--Yang edge singularities, functional methods, and Taylor or contour-based expansions around $\mu_B = 0$ \cite{Borsanyi:2021sxv,Noronha-Hostler:2019ayj,Kahangirwe:2024cny, Abuali:2025tbd, Vovchenko:2017gkg,Borsanyi:2020fev,Bollweg:2022rps,Borsanyi:2025dyp, Fu:2019hdw,Gunkel:2021oya,Gao:2020fbl,Hippert:2023bel,Basar:2023nkp,Clarke:2024ugt,Shah:2024img,Adam:2025phc}. Extrapolations of lattice QCD data from zero baryon density generally indicate that the crossover transition persists at least up to $\mu_B \lesssim 450~\text{MeV}$ \cite{Vovchenko:2017gkg,Borsanyi:2020fev,Bollweg:2022rps,Borsanyi:2025dyp}. 
More recently, however, a number of QCD-motivated effective models, lattice-based extrapolation schemes, and theoretical analyses of heavy-ion data have suggested the possibility of a critical point in the approximate range $450~\text{MeV} < \mu_{B,c} < 700~\text{MeV}$ \cite{Fu:2019hdw,Gunkel:2021oya,Gao:2020fbl,Hippert:2023bel,Basar:2023nkp,Clarke:2024ugt,Shah:2024img,Adam:2025phc,Sorensen:2024mry}.

Recently, we introduced a novel method based on the expansion of constant entropy density contours from zero $\mu_B$ \cite{Shah:2024img}. The method was used to predict the CP location, using lattice QCD results and truncating the expansion at order $\mathcal{O}(\mu_B^2)$. It was then employed by the Wuppertal Budapest LQCD collaboration under strangeness neutral conditions, to predict that the CP is excluded for $\mu_B <450 $ MeV \cite{Borsanyi:2024xrx}. The method was also explored in Ref. \cite{Marczenko:2025znt} in the context of its possible limitations in correctly predicting the critical point. 

In this work, we present a systematic investigation of the method's performance within various models for the QCD equation of state, in particular, its reliability and accuracy for determining the CP.
To this end, we apply it to  approaches both with and without a critical point. 
These approaches include an ideal gas of massless quarks and gluons, hadron resonance gas (HRG) models \cite{Vovchenko:2014pka,Vovchenko:2015xja,Vovchenko:2017xad} with various implementations of hadronic interactions, the cluster expansion model (CEM) \cite{Vovchenko:2017gkg}, the Nambu--Jona-Lasinio (NJL) model \cite{Buballa:2003qv}, the holographic Einstein-Maxwell-dilaton (EMD) model \cite{Hippert:2023bel}, and the data from the functional renormalization group approach \cite{Lu:2025cls}.
We also consider expansions of enthalpy and energy density contours in lieu of entropy density, which produce consistent results.
We find that the method produces an accurate (to 5\% accuracy) estimate for the CP in EMD and DSE/fRG approaches, both of which are consistent with lattice QCD data at $\mu_B = 0$.
However, we also find a spurious CP at $(T,\mu_B) \sim (90,800)$~MeV in models that incorporate hadron-resonance degrees of freedom. These contours start at a temperature $T_0 < 120$ MeV at $\mu_B = 0$ and indicate that the method cannot be used in that regime.

Finally, we also expand our previous analysis of lattice QCD results~\cite{Shah:2024img}.
In particular, we reconstruct the entire equation of state by integrating the entropy density to obtain the pressure as a function of $T$ and $\mu_B$, its natural variables.
We then analyze the behavior of isentropic trajectories along the phase diagram, especially in the vicinity of the CP.
We also extract Lee-Yang edge singularities and study their evolution in the complex $\mu_B$ and how they are connected to the CP singularity on the real $\mu_B$ axis, and to the Roberge-Weiss phase transition at imaginary $\mu_B$.

Our manuscript is organized as follows: In Section~\ref{sec:method}, we introduce the method of expanding constant entropy density contours. In Section~\ref{sec:applications}, we apply this method to various effective models: we begin with an ideal gas of massless quarks and gluons in Subsec.~\ref{sec:massless}, followed by the Hadron Resonance Gas model and its variants, including the Cluster Expansion Model, in Subsec.~\ref{sec:hrg}. We then explore the Nambu--Jona-Lasinio model in Subsec.~\ref{sec:njl}, the Holographic Einstein-Maxwell-Dilaton model in Subsec.~\ref{sec:holography}, and functional QCD approaches in Subsec.~\ref{sec:frg}. Finally, Section~\ref{sec:lattice} presents the construction of a lattice-based equation of state and its properties, followed by our conclusions in Section~\ref{sec:conclusions}.

\section{Contours of constant entropy density method} \label{sec:method}
\begin{figure*}[t]
    \includegraphics[width=\linewidth]{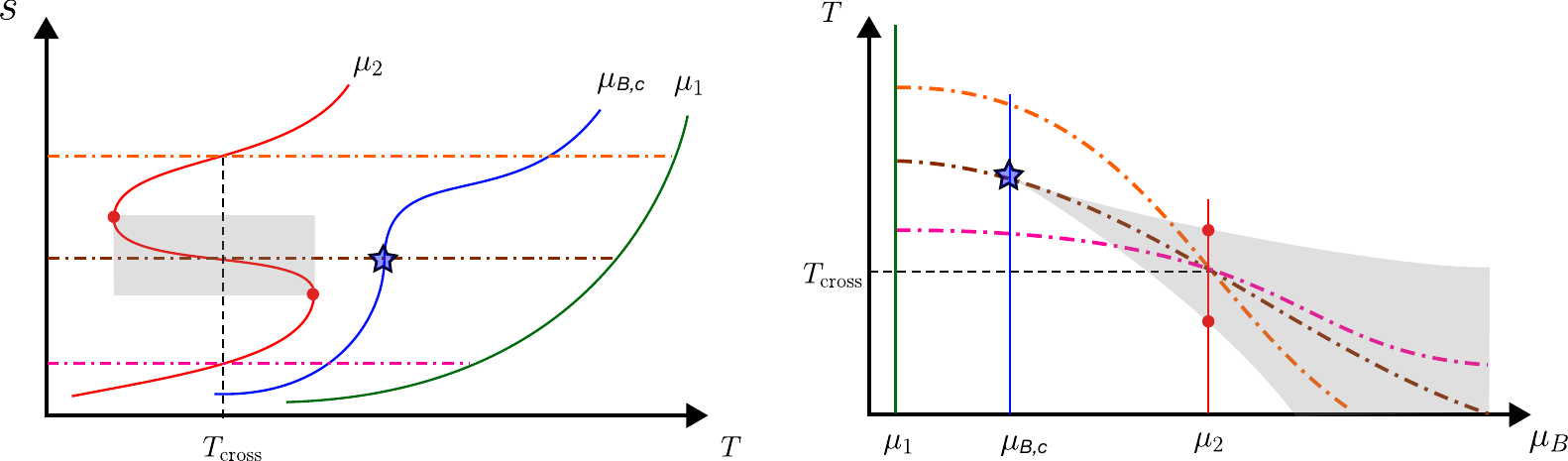}

   \caption{ \justifying \small 
\textit{Left panel:} Entropy as a function of temperature for three representative baryon chemical potentials, with $\mu_1 < \mu_{B,c} < \mu_2$. 
\textit{Right panel:} The corresponding constant-entropy trajectories in the $(T, \mu_B)$ plane. 
The blue star marks the location of the critical point, the shaded region indicates the spinodal domain, and the red dots highlight the spinodal points at $\mu_B = \mu_2$.  Figure taken from \cite{Shah:2024img}
}

    \label{fig:cartoon_firstorder}
\end{figure*}

The idea of developing an expansion scheme to search for a critical point by following the contours of constant entropy density was first developed in \cite{Shah:2024img}. The entropy density is used because, at the mean-field level, it is a multi-valued function in the presence of a first order phase transition. This implies that, if the Maxwell construction is not performed to obtain the most stable solution, one will observe three values for the entropy, corresponding to the stable, metastable and unstable branches (spinodal region) as illustrated in Fig. \ref{fig:cartoon_firstorder}

The left panel of Fig. \ref{fig:cartoon_firstorder} depicts curves for the entropy density $s$ as a function of the temperature $T$ for different values of the chemical potential $\mu_B$. At low chemical potential $\mu_1$, there is a crossover as $s$ increases monotonically with  $T$; at a higher chemical potential $\mu_{B,c}$, a divergent slope in $s$ at  $T=T_c$ signals the critical point; for $\mu_B=\mu_2>\mu_{B,c}$, three values of entropy can be obtained for the same $(T,\mu_B)$ pair, implying that, when we map this to a $(T,\mu_B)$ plane, the entropy lines should cross as shown on the right panel of Fig. \ref{fig:cartoon_firstorder}. The crossings span through the entire spinodal region of the first order phase transition. 

Since lattice QCD simulations are feasible only at vanishing chemical potential, we need to use an expansion to obtain the constant entropy density contours:
\begin{equation}
    T_s(\mu_B; T_0) \approx T_0 + \sum_{n=1}^N \coef_{2n}(T_0) \,\frac{\mu_B^{2n}}{(2\,n)!} + \mathcal{O}\left(\mu_B^{2(N+1)}\right),
    \label{eq:expSsup}
\end{equation}
where the expansion coefficients are evaluated at fixed $s$,
\begin{equation}
\coef_{2n}(T_0) = \left. \left(\frac{\partial^{2n} T}{\partial \mu_B^{2n}}\right)_s \right|_{T = T_0, \mu_B = 0}.
\end{equation}
Due to the fact that $\rho_B = 0$ for all temperatures at $\mu_B = 0$, 
all odd power coefficients are zero when starting the expansion from $\mu_B = 0$, so the leading-order coefficient becomes:
\begin{equation}
\label{eq:app:C2}
\coef_2(T_0) = \left. \left(\frac{\partial^2 T}{\partial \mu_B^2}\right)_s \right|_{T=T_0}^{\mu_B = 0} = -\frac{2T_0 \chi_2^{B}(T_0) + T_0^2 \chi_2^{B'}(T_0)}{s'(T_0)}.
\end{equation}
At $\mathcal{O}(\mu_B^2)$, the function $T_s(\mu_B;T_0)$ reads
\begin{equation}
\label{eq:TsO2}
T_s(\mu_B;T_0) = T_0 + \coef_{2}(T_0) \frac{\mu_B^2}{2}.
\end{equation}

Using the fact that the critical point is an inflection point i.e. $\frac{\partial^2 T}{\partial s^2} = 0$, and that the entropy has a divergent slope at the CP so that $\frac{\partial T}{\partial s} = 0$, the critical point from the expansion can be determined by applying this pair of equations to the expansion, hence finding $\left(\frac{\partial T_s}{\partial T_0}\right)_{\mu_B} = 0$ and $\left(\frac{\partial^2 T_s}{\partial T_0^2}\right)_{\mu_B} = 0$. This is due to the fact that $\left(\frac{\partial s}{\partial T_0}\right)_{\mu_B=0}$ is a positive monotonically increasing function and is not equal to zero at any finite temperature $T_0$.
Denoting $\mu_{B,c}$ and $T_{0,c}$ as the values of $\mu_B$ and $T_0$ corresponding to the CP, the first equation, $\left(\frac{\partial T_s}{\partial T_0}\right)_{\mu_B} = 0$, yields the relationship between $\mu_{B,c}$ and $T_{0,c}$:
\begin{equation}
\label{eq:spinodal}
1 + \coef_2'(T_{0,c}) \frac{\mu_{B,c}^2}{2} = 0 \quad \Rightarrow \quad \mu_{B,c} = \sqrt{-\frac{2}{\coef_2'(T_{0,c})}}. 
\end{equation}
This equation also determines the spinodal lines at $\mu_B > \mu_{B,c}$ where it has two solutions with respect to $T_0$.

The second equation, $\left(\frac{\partial^2 T_s}{\partial T_0^2}\right)_{\mu_B} = 0$, corresponds to
\begin{equation}
\label{eq:T0c}
\coef_2''(T_{0,c}) = 0,
\end{equation}
which determines $T_{0,c}$.
The determination of the CP location thus proceeds by solving Eq.~\eqref{eq:T0c} for $T_{0,c}$, plugging the result into Eq.~\eqref{eq:spinodal} to determine $\mu_{B,c}$, and  computing $T_c = T_s(\mu_{B,c};T_{0,c})$ through Eq.~\eqref{eq:TsO2}. At $\mu_B > \mu_{B,c}$, the Maxwell construction of equal areas for entropy density can be utilized to determine the phase coexistence line.

One can also find an expression for the fourth order coefficient
\begin{multline}
    \alpha_4(T_0) 
    = \left.\frac{\partial^4 T}{\partial\mu_B^4}\right|_s =-\frac{\chi_4'(T_0)}{s'(T_0)} - \frac{3 s''(T_0) \alpha_2^2(T_0)}{s'(T_0)} - \\
     \frac{6 \alpha_2(T_0) (2\chi_2(T_0) + 4T\chi_2'(T_0)+T^2\chi_2''(T_0))}{s'(T_0)} .
\end{multline}
Applying the expression above to lattice QCD results is quite challenging, as one would also require second order derivatives of $\alpha_4$ to locate the CP, and extracting higher order temperature derivatives from discrete points leads to a larger uncertainty in the CP location. Due to this, estimating the radius of convergence of the expansion becomes considerably more complicated as the uncertainties required in the input grows significantly when one performs higher ordered terms in the expansion. Hence, in this work we study the behavior of the contours truncating the series so that they are parabolic as Ref.~\cite{Borsanyi:2025dyp} suggests that constant entropy contours in the $T-\mu_B^2$ plane are approximately linear in the imaginary chemical potential regime, supporting the use of the $\mathcal{O}(\mu_B^2)$ expansion.
\smallskip

\textbf{Energy density contours:} A similar extrapolation can be performed on contours of constant energy density, as this quantity should also have multiple solutions in the presence of a first order phase transition and, thus, a corresponding S-shaped behavior. The energy density is defined as:
\begin{equation}
    \epsilon (T,\mu_B) = -P + T s + \mu_B n_B.
    \label{eq:edens}
\end{equation}
When we apply our method to the case of energy density, namely finding the contours where $d\epsilon = 0$, we obtain this expression for the coefficient of expansion to second order:
\begin{equation}
    \label{eq:alpha_2_energy}
	\alpha_2^{\epsilon} (T_0) = \alpha_2 (T_0) - \frac{ T_0^2 \chi_2^B (T_0)}{T_0s'(T_0)}.
\end{equation}
Here, $\alpha_2(T_0)$ is the coefficient of expansion of the constant entropy density contours as given in Eq. \eqref{eq:app:C2}.
Due to the fact that the energy density is also a multi-valued observable at the first order phase transition, %
similar conditions as the one found for the entropy density contours apply here as well, namely 
$(\frac{d\epsilon}{dT})_{\mu_B} = 0$ and $(\frac{d^2\epsilon}{dT^2})_{\mu_B} = 0$.
\smallskip
They lead to the following equations in terms of $\alpha_2^{\epsilon}(T_0)$: $\alpha_2^{\epsilon''}(T_0)=0$ and $\mu_{B,c}= \sqrt{\frac{-2}{\alpha_2^{\epsilon'}(T_{0c})}}$.

\textbf{Enthalpy density contours:} Another quantity for which our method can be used is the enthalpy density, which is defined in the grand canonical ensemble as: 
\begin{equation}
    h (T, \mu_B) = T s + \mu_B  n_B.
    \label{eq:hdens}
\end{equation}
We find the following expression for the second order expansion coefficient of the constant enthalpy density contours 

\begin{equation}
    \label{eq:alpha_2_enthalpy}
    \alpha_2^h (T_0) =
    - \frac{T_0^2 \left[T_0 \chi_2^{B'}(T_0) + 4\,\chi_2^B (T_0)\right]}{T_0 s' (T_0) + s (T_0)}.
\end{equation}
Similar conditions relating to the inflection point and the divergent slope of the enthalpy density at the CP can be applied to obtain the critical point, $(\frac{dh}{dT})_{\mu_B} = 0$ and $(\frac{d^2h}{dT^2})_{\mu_B} = 0$. They yield the equations $\alpha_2^{h''}(T_0)=0$ and $\mu_{B,c}= \sqrt{\frac{-2}{\alpha_2^{h'}(T_{0c})}}$.

While contours of constant $s$, $\epsilon$ and $h$ correspond to different expansion coefficients, all of them should lead to the same location for the critical point, up to truncation errors. 
That is, any differences between the predictions extracted from these three quantities must be attributed to the truncation of the corresponding expansions in $\mu_B$. 
This might provide us with a way to gauge the robustness of predictions regarding the  existence and location of the critical point, namely by comparing results from expansions at constant  entropy, energy and enthalpy densities. However, conclusions should be taken with care, since the errors for each of these expansions may not be entirely independent, and could be correlated to some extent. 
As we show below, while the agreement among different observables constitutes a necessary condition for the reliable estimation of the CP, taken alone, it does not constitute a sufficient condition.
In the following, it will also be instructive to explore which of these expansions yields the best prediction within each of the models we analyze. 
\section{Applications} \label{sec:applications}
\subsection{Ideal gas of massless quarks and gluons} \label{sec:massless}
The first model we test is the limit where the particles in QCD are assumed to be massless quarks and gluons. This approximation is well-motivated for exploratory and comparative studies, where the effects of finite masses and interactions are considered subdominant. Due to the fact that the particles are treated as a non-interacting gas, there is no phase transition in the system. This constitutes a first check to see if the expansion finds any spurious crossings despite the absence of a CP. The total pressure in this model receives contributions from massless gluons and $N_f$ flavors of massless quarks given as follows~\cite{Vovchenko:2017xad}:
\begin{equation}
\frac{P}{T^4} = \frac{8\pi^2}{45} + \sum_{f=u,d,s} \left[ \frac{7\pi^2}{60} + \frac{1}{2} \left( \frac{\mu_f}{T} \right)^2 + \frac{1}{4\pi^2} \left( \frac{\mu_f}{T} \right)^4 \right]
\label{eq:SB_pressure}
\end{equation}
where the first term represents the contribution from gluons, and the sum captures the quark sector. This expression is the Stefan–Boltzmann pressure of a massless ideal gas of quarks and anti-quarks at finite chemical potential.

Assuming vanishing strange and electric charge chemical potentials ($\mu_S = \mu_Q = 0$), and identifying the quark chemical potentials as $\mu_f = \mu_B/3$, the pressure normalized by $T^4$ takes the form: 
\begin{equation}
    \frac{P}{T^4} =  \frac{8\pi^2}{45} + \left[ \frac{7\pi^2}{20} + \frac{3}{2} \left( \frac{\mu_B}{3T} \right)^2 + \frac{3}{4\pi^2} \left( \frac{\mu_B}{3T} \right)^4 \right].
\end{equation}
The baryon susceptibility defined as $\chi_2^B = \frac{\partial^2 P/T^4}{\partial (\mu_B/T)^2}$ becomes:  
\begin{equation}
    \chi_2^B = \frac{1}{3} + \frac{\mu_B^2}{(3\pi T)^2}.
\end{equation}
Similarly, the entropy density in the model is given as:
\begin{equation}
    s = \frac{\partial P}{\partial T} = 4T^3 C + T\mu_B^2/3, \qquad C = \frac{8\pi^2}{45} + \frac{7\pi^2}{20}.
\end{equation}

Now that the entropy density and the second-order baryonic susceptibility are calculated, one can use them to get the coefficient of expansion $\alpha_2$ using Eq. \eqref{eq:app:C2}:
\begin{equation}
    \alpha_2(T_0) = \frac{-1}{18TC} \implies \alpha_2''(T_0) = \frac{-1}{9T^3C}.
\end{equation}
From the equation above, one can see that $\alpha_2''$ goes to zero at infinity, only asymptotically satisfying the condition for determining the critical $T_{0c}$ from the expansion, implying that no spurious phase transition is found at finite $T$. 

\begin{figure}[h]
    \centering
    \includegraphics[width=0.9\linewidth]{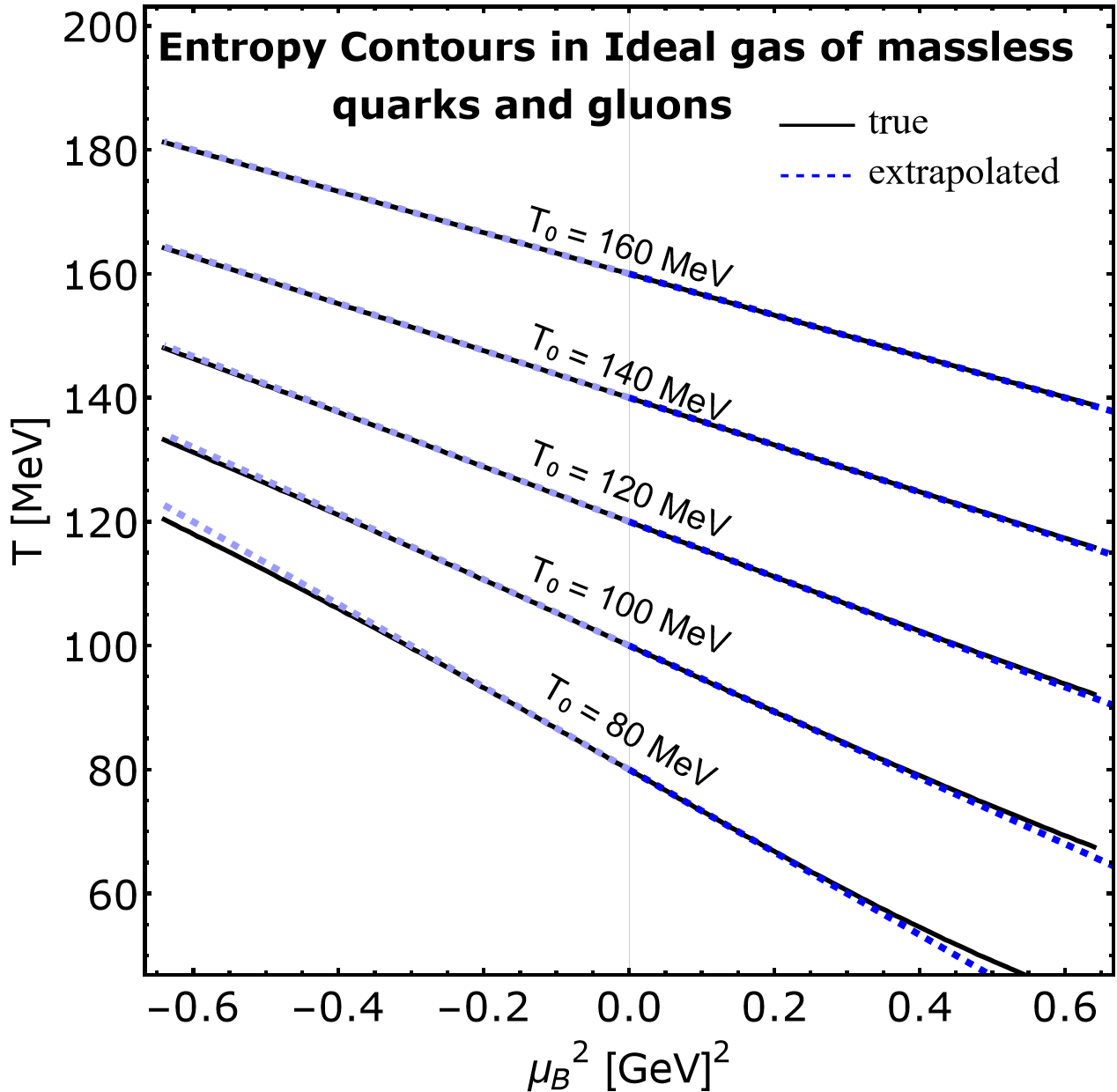}

    \caption{\justifying True (black, solid) vs truncated (blue, dashed) contours starting from different temperatures at zero $\mu_B$ for an ideal gas of massless quarks and gluons for both imaginary and real $\mu_B$. }
    \label{fig:q&g_contours}
\end{figure}

Comparing the actual  constant entropy contours of the model against the contours obtained from the extrapolation, see  Fig.~\ref{fig:q&g_contours}, we find that the extrapolated contours are consistent with the contours of the full model  both for real and imaginary baryon chemical potential. 
We confirm that, when calculating the contours based on the truncated Taylor expansion, no artificial critical behavior appears in the imaginary plane and the contours never cross. 
One can similarly calculate the coefficient of expansion for constant energy density and enthalpy density contours to find that their second temperature derivatives also approach zero only as $T$ approaches infinity.

Notice that, at least in this simple example, one would be able to anticipate the good agreement between the quadratic extrapolation and the full model contours at real values of $\mu_B$ by checking the agreement at imaginary values of $\mu_B$.

\vspace{-1cm}
\subsection{Hadronic Models} \label{sec:hrg}

\subsubsection{Ideal Hadron Resonance Gas Model} \label{sec:ideal_hrg}
A popular model which describes the confined phase of strongly interacting matter is the ideal Hadron Resonance Gas (HRG) model, which matches the results for lattice QCD thermodynamics at low temperatures \cite{Vovchenko:2014pka}. This model was originally developed in \cite{Hagedorn:1967tlw}, following the idea that resonance formation dominates the interactions between ground-state hadrons in the confined phase. The most standard case is the ideal HRG model, which treats the particles in the system as point-like and non-interacting. The influence of interactions is therefore represented indirectly through the inclusion of all known resonant states in the hadron spectrum. Within the Grand Canonical Ensemble (GCE), the pressure of the HRG model is given as a sum over partial pressures of all hadron species,
\begin{eqnarray}
    \label{eq:pid}
    P^{id}_i (T,\mu) &=& \frac{d}{3}\int \frac{d^3k}{(2\pi)^3}\frac{\textbf{k}^2}{\sqrt{m_i^2 + \textbf{k}^2}} \times
    \\
    &&\times\left[ \exp \left(\frac{\sqrt{m_i^2 + \textbf{k}^2}-\mu_i }{T} \right)  + \eta\right]^{-1}.
    \nonumber
\end{eqnarray}
In Eq. (\ref{eq:pid}), $\eta$ is $+1$ for Fermi statistics, $-1$ for Bose statistics and 0 for the Boltzmann approximation. 
The chemical potential for each species is given by $\mu_i = B_i \mu_B + Q_i \mu_Q + S_i \mu_S$.
In this work, we use the PDG 2021+ hadron list based on Data Tables provided in Ref.~\cite{SanMartin:2023zhv} and use the \texttt{Thermal-FIST} package~\cite{Vovchenko:2019pjl} for calculating HRG thermodynamics.
\begin{figure}
    \centering
    \includegraphics[width=0.9\linewidth]{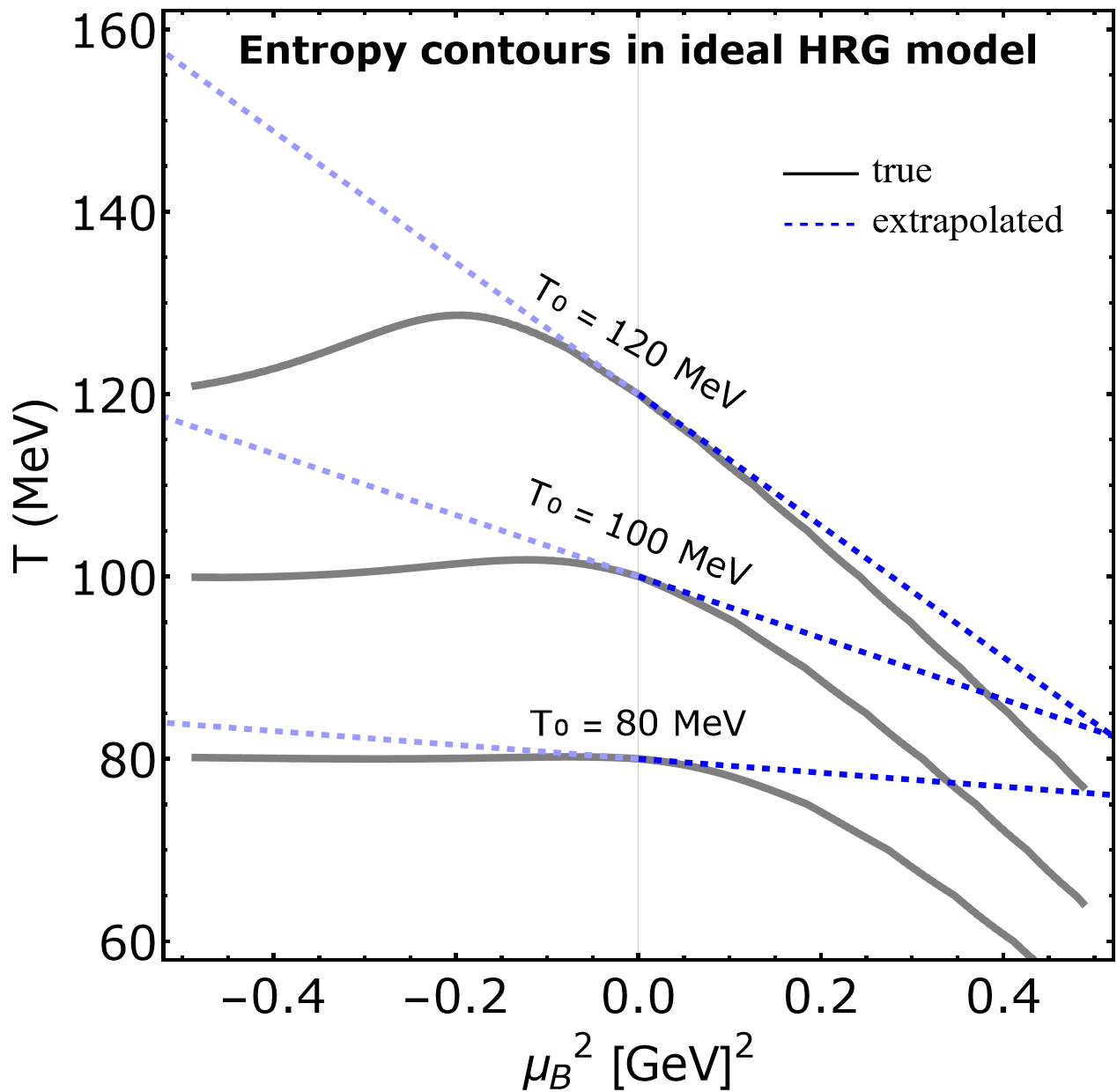}
    \caption{\justifying True contours (black, solid) shown against the extrapolated contours (blue, dashed) in an ideal hadron resonance gas model following Boltzmann statistics in the real and imaginary plane. The discrepancy in the extrapolated contours displays the invalidity of the parabolic approximation for the expansion in this case.}
        \label{fig:IHRG}
\end{figure}

\begin{figure*}[bth]
    \centering
    \includegraphics[width=0.335\textwidth]{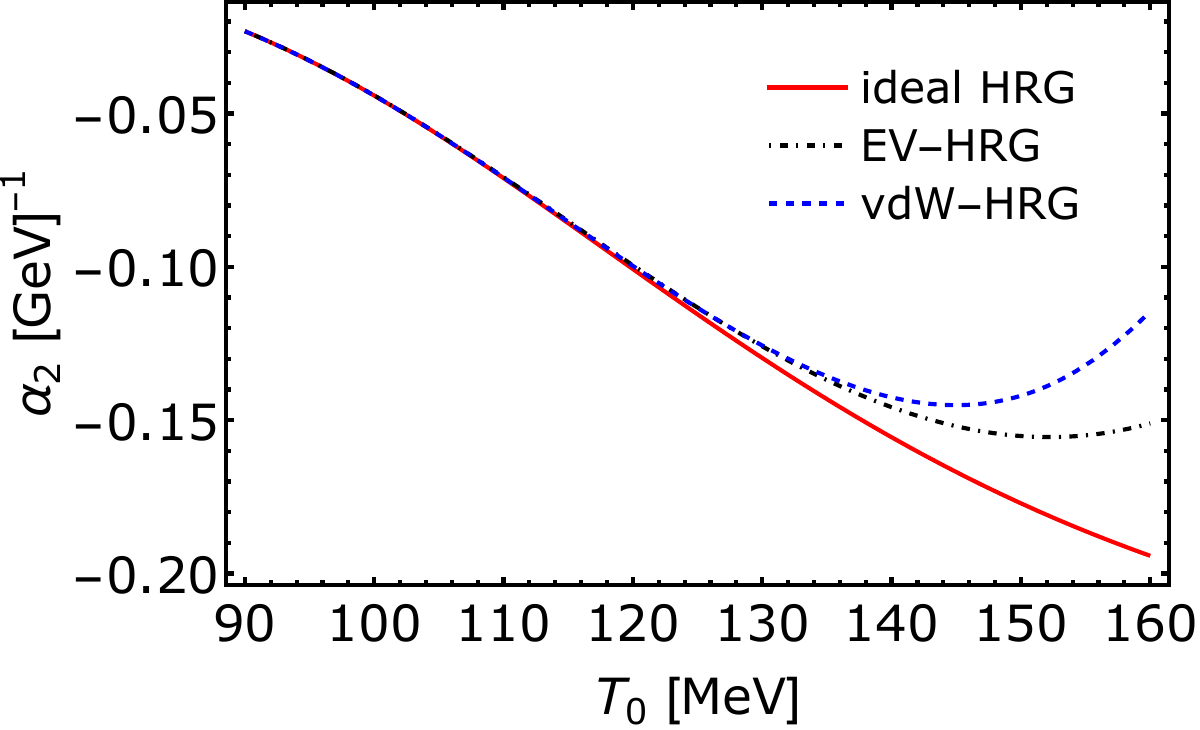}
    \includegraphics[width=0.32\linewidth]{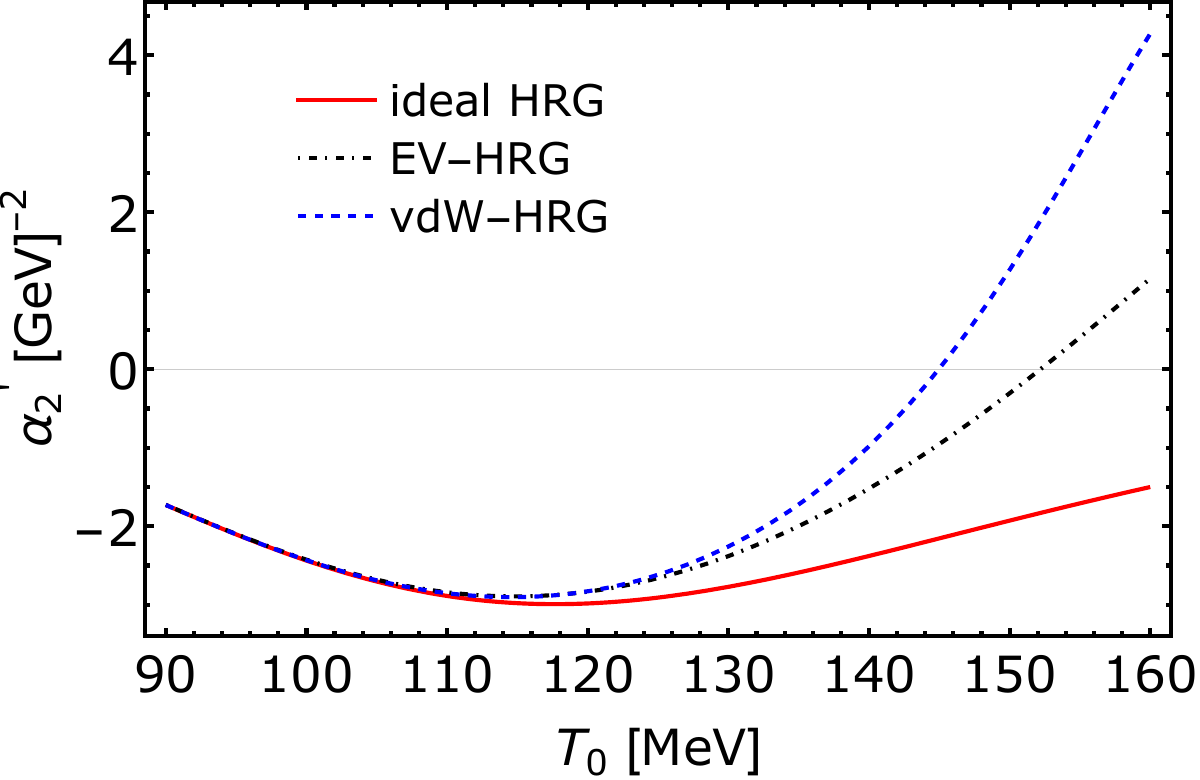}\includegraphics[width=0.33\linewidth]{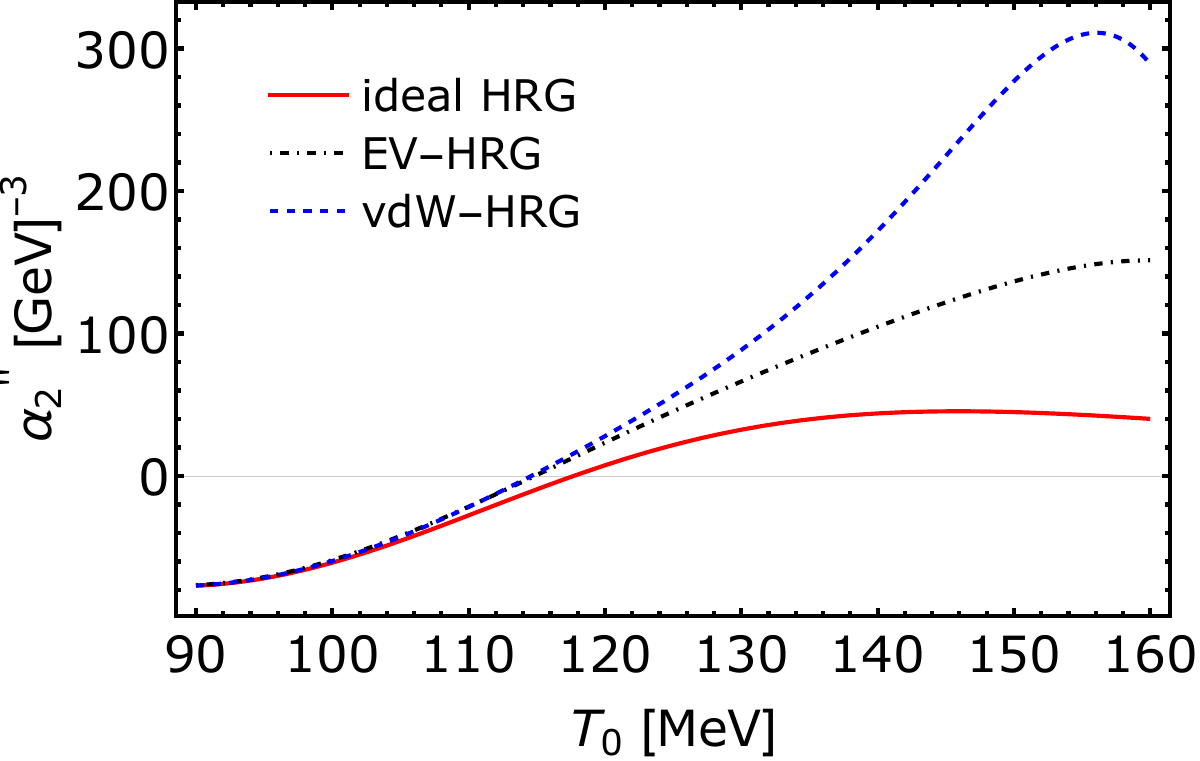}
    \caption{\justifying \small Constant entropy contour expansion coefficient $\alpha_2$ and its temperature derivatives across various HRG-based models as functions of the temperature at vanishing $\mu_B$.}
    \label{fig:alphas_HRG}
\end{figure*}

We calculate the coefficient of expansion $\alpha_2$ using Eq.~(\ref{eq:app:C2}), together with its temperature derivatives as shown in Fig.~\ref{fig:alphas_HRG}. For the ideal HRG model, $\alpha_2''$ crosses zero, which yields a \emph{spurious} 
critical-point solution via the contour criteria in Eqs.~\eqref{eq:spinodal} and \eqref{eq:T0c}. This would correspond to $(T,\mu_B) = (86,817)$ MeV from entropy-density contours. Since the ideal HRG model has no phase transition, this should instead be interpreted as the onset of non-parabolic contour behavior and, therefore, the breakdown scale of the quadratic truncation. Consistently, the same procedure applied to other observables gives similarly spurious results: $(T,\mu_B) = (88,791)$ MeV from enthalpy density and $(86,777)$ MeV from energy density.
It was shown in \cite{Marczenko:2025znt} that this fake critical point is due to the mesonic degrees of freedom dominating at zero baryon chemical potential and high temperature. 

We now take a similar approach to that used in the previous section for the ideal gas of massless quarks and gluons, and analyze the contours of constant entropy density in the ideal hadron resonance gas (HRG) model. To this end, we consider the ideal HRG with Boltzmann statistics in both the real and imaginary chemical potential planes. The Boltzmann approximation is chosen because it represents the simplest version of the ideal HRG, and other variants of the model are expected to exhibit quantitatively similar contour behavior. 

If possible, one should study the behavior of the contours in the imaginary $\mu_B$ plane. If the quadratic expansion works in the imaginary plane, then the expansion in the real plane is expected to be trustworthy. For this model, in Fig. \ref{fig:IHRG} we compare the full entropy density contours obtained directly from the model with those reconstructed from the truncated expansion at order $\mu_B^2$, both for real and imaginary $\mu_B$. 
 From Fig.~\ref{fig:IHRG}, it can be seen that the quadratic approximation fails to reproduce the contours adequately in both planes for this model. In the real plane, the contours that start from low $T_0$ at $\mu_B=0$ are flat when expanded to higher $\mu_B$ values at order $\mu_B^2$, while the true contours of the model have an exponential decline when $\mu_B$ is close to the mass of the proton. However, the contours that start from high temperatures suggest a quadratic behavior and the truncated contours represent the full ones accurately. Due to the quadratic term not being able to reproduce the contours at low temperature, the fake crossings appear, giving rise to the fake critical point and first order phase transition in the real $\mu_B$ plane.

\subsubsection{Excluded Volume HRG model} \label{sec:ev_hrg}

In the excluded-volume (EV) formulation of the HRG model, one adds a short-range repulsive interaction akin to hard-sphere repulsion~\cite{Rischke:1991ke}.
In this model, the available volume in the system for a hadron to move is effectively reduced by repulsion.
Typically, repulsive interactions are introduced for baryon-baryon and antibaryon-antibaryon pairs~\cite{Satarov:2016peb,Vovchenko:2016rkn} and this is the approach we employ here.
In this case, the pressure is separated into contributions from non-interacting mesons, and interacting baryons and antibaryons:
\begin{equation}
P^{\rm ev}(T,\mu) = P^{\rm id}_M(T,\mu) + P^{\rm ev}_B(T,\mu) + P^{\rm ev}_{\bar{B}}(T,\mu),
\end{equation}
where the partial pressures of (anti)baryons are determined from the equations
\begin{align}
P^{\rm ev}_{B(\bar{B})}(T,\mu) & = \sum_{i \in B(\bar{B})} P^{\rm id}_i(T,\mu_i^*), \\
\mu_i^* & = \mu_i - b \,P^{\rm ev}_{B(\bar{B})}(T,\mu).
\end{align}
Therefore, to solve the excluded volume model in the grand-canonical ensemble, one needs to obtain the solution of the above transcendental equations for each point of the $(T,\mu_B)$ plane. For our hadron resonance gas model, the excluded volume parameter for (anti)baryons is taken to be $b=1$ fm$^3$. 
This value of $b$ is chosen based on the fit to lattice results for the susceptibilities at $\mu_B=0$~\cite{Vovchenko:2016rkn, Vovchenko:2017xad,Karthein:2021cmb}. 
We use \texttt{Thermal-FIST}~\cite{Vovchenko:2019pjl} for calculations.
\begin{figure}[bth]
    \centering
    \includegraphics[width=0.45\textwidth]{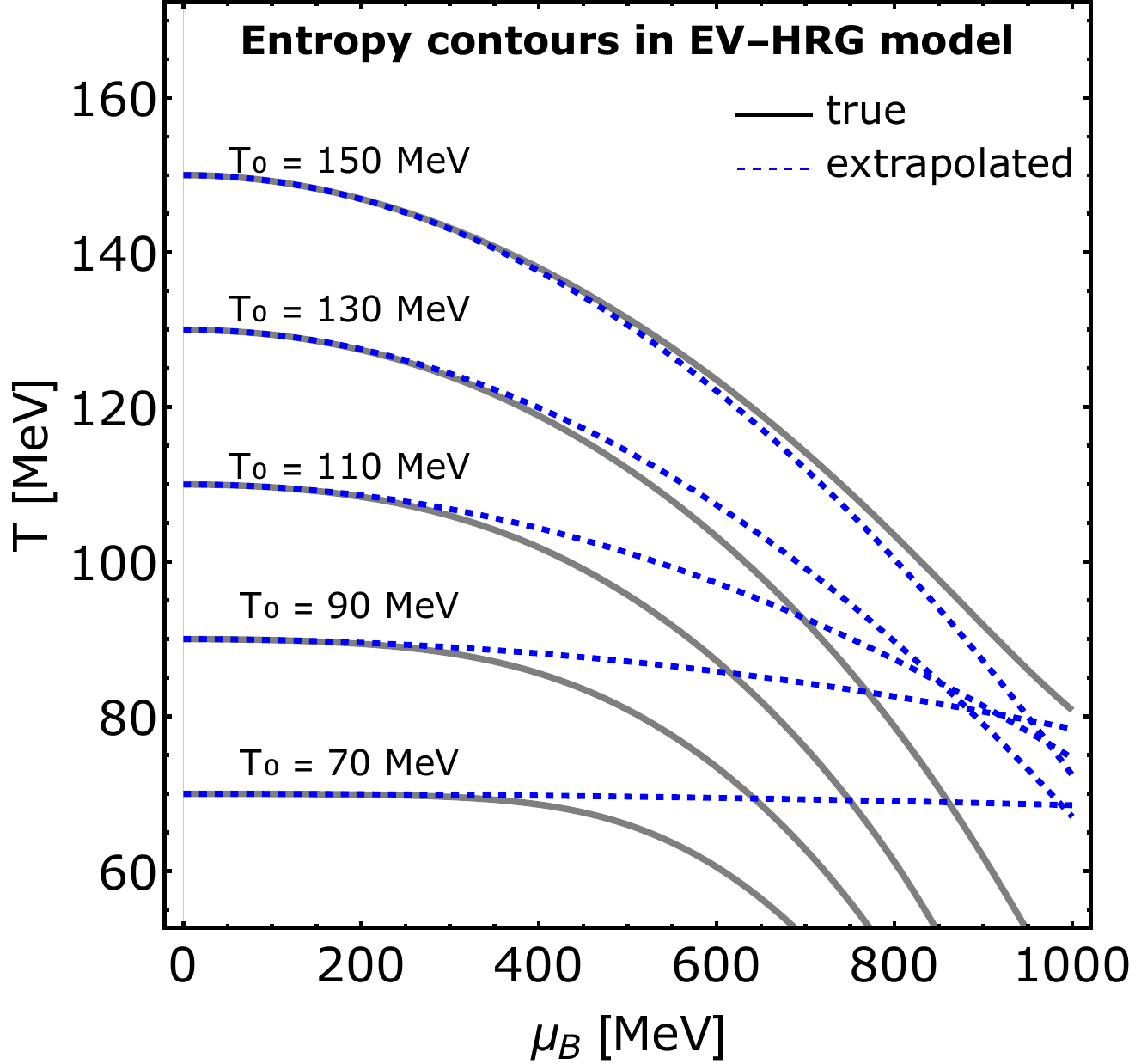}
    \caption{\justifying \small Full (solid,black) vs truncated (blue,dashed) contours of constant entropy density in the excluded volume HRG model in the real baryon chemical potential plane, starting from different initial temperatures.}
    \label{fig:cont_compare_EV}
\end{figure}

When one uses the excluded volume HRG, which describes the lattice QCD results similarly to the ideal HRG at vanishing $\mu_B$, and calculates the constant entropy density contours, a similar behavior for the truncated contours is found. As seen in Fig. \ref{fig:cont_compare_EV}, the contours truncated at the quadratic order deviate from the actual contours when they start from lower $T_0$ at $\mu_B=0$. The contours at higher temperature still look quadratic and are able to reproduce the actual contours up to higher chemical potential. Even so, as the actual contours are not reproduced properly at low temperature and high $\mu_B$, a fake critical point is found due to the crossing of the contours. The unphysical critical point is obtained at $(T,\mu_B)$ = $(85,831)$ MeV.

From the results in the ideal and excluded volume HRG, we conclude that the expansion up to second order in $\mu_B$ is not able to describe the system at low temperature, namely in the hadronic phase. The truncation of the contours at $\mathcal{O}(\mu_B^2)$ hints at a critical point at $\mu_B>$ 800 MeV, suggesting that such high values of $\mu_B$ cannot be probed using this expansion. 

\subsubsection{Quantum van der Waals HRG model} \label{sec:vdw_hrg}
In the QvdW-HRG model, both attraction and repulsion are incorporated, and the parameters are fixed to reproduce the saturation density and binding energy of nuclear matter \cite{Vovchenko:2014pka, Vovchenko:2016rkn}. 
Just like for the case of the EV-HRG model, the pressure contains contributions from non-interacting mesons, interacting baryons and antibaryons. 
We do not incorporate attraction in meson-meson or meson-baryon channels, since such interactions are already incorporated in the HRG model to a large extent by including resonances in the particle list.
The attractive and repulsive  interactions are assumed to be the same between all pairs of baryons and antibaryons in the list of hadrons and resonances. 
The presence of the attraction terms modifies the transcendental equations slightly.%
\footnote{Note that for $a = 0$ the QvdW-HRG model reduces to the EV-HRG model.} 
Namely, one has~\cite{Vovchenko:2017zpj}
\begin{align}
P^{\rm vdW}_{B(\bar{B})}(T,\mu) & = \sum_{i \in B(\bar{B})} P^{\rm id}_i(T,\mu_i^*) - a \,n_{B(\bar{B})}^2, \\
n_{B(\bar{B})}(T,\mu) & = \frac{n_{B(\bar{B})^{\rm id}}(T,\mu^*)}{1 + b \, n_{B(\bar{B})^{\rm id}}(T,\mu^*)}, 
\end{align}
\begin{equation}
    \mu_i^* = \mu_i - b\, P^{\rm vdW}_{B(\bar{B})}(T,\mu) - a b \, n_{B(\bar{B})}^2 + 2 a \,n_{B(\bar{B})},
\end{equation}
where $n_{B(\bar{B})}(T,\mu)$ is the number density of (anti)baryons in the van der Waals model, while $n_{B(\bar{B})^{\rm id}}(T,\mu^*)$ is the ideal gas density evaluated at shifted chemical potentials $\mu_i^*$.
For our work, we choose $a = 329$ MeV fm$^3$ and $b=3.42$ fm$^3$, as these values provide the correct description of the nuclear saturation density and the binding energy of the nucleons~\cite{Vovchenko:2015vxa}. These values also yield a fair description of the lattice QCD thermodynamics at vanishing chemical potential~\cite{Vovchenko:2016rkn}. 
Also in this case we use \texttt{Thermal-FIST}~\cite{Vovchenko:2019pjl} for calculations.
When the coefficient of expansion for the constant entropy density contours is obtained from the zero density equation of state, as seen in Fig.~\ref{fig:alphas_HRG}, the quantity $\alpha_2''$ goes to zero at a critical value of $T_{0c}=115$ MeV, similar to the values found in the other HRG-based models. This similarity arises due to the mesonic degrees of freedom being dominant  at vanishing baryon chemical potential and temperatures lower than the pion mass, whereas the interactions amongst baryons become subdominant, leading to all the HRG models having a similar behavior in this regime. 
It is also observed that the fake critical points obtained in HRG-based models are consistently found at values of $T_0$ below 120 MeV. 
In this particular model, the critical point location using the extrapolation of $s$-contours is found at $(T,\mu_B) \approx (85,830)$ MeV, which is very close to the one obtained in the excluded volume HRG model. This suggests that, even though there is a nuclear liquid-gas phase transition in the model, the coefficient $\alpha_2$ at zero chemical potential is not particularly sensitive to it and provides the same fake crossings of entropy density contours. For energy density contours, the CP from extrapolation is obtained at $(T,\mu_B) \approx (86,789)$ MeV while for enthalpy density contours, its location is $(T,\mu_B) \approx (87,805)$ MeV. Thus, also for this version of the HRG model, the extrapolation of these contours to order $\mu_B^2$ is insufficient. 

We note that the QvdW-HRG model describes a nuclear liquid-gas transition and does contain a CP at $(T,\mu_B) \simeq (20, 908)$~MeV.
In principle, the CP at $(T,\mu_B) \approx (85,830)$ could have been related to the nuclear CP, but given the similarity of this result to those obtained in the ideal HRG and EV-HRG cases, we consider it to be a spurious one.
The nuclear liquid-gas CP is located at values of $\mu_B$ which are too high (and values of $T$ which are too low, giving $\mu_B/T \sim 45$) to be captured by a parabolic extrapolation from $\mu_B = 0$.

Having established that HRG-based descriptions with both ideal and interacting systems generate spurious critical points when truncated at $\mathcal{O} (\mu_B^2)$, we explored the expansion in another lattice-constrained hadronic description of QCD by using the Cluster Expansion Model (CEM)~\cite{Vovchenko:2017gkg}. 
The CEM provides an equation of state at finite baryon density based on lattice QCD, obtained from an expansion in baryonic fugacities \cite{Vovchenko:2017xad,Vovchenko:2018zgt,Bellwied:2015lba}. The equation of state at real $\mu_B$ does not incorporate a first order phase transition. We used the parametrized version of the CEM \cite{Vovchenko:2019vsm} to obtain the coefficient of the expansion for the constant entropy contours. We found that the 2nd derivative of the expansion coefficient vanishes at a similar temperature to that of the HRG models. 
For this reason, we obtain a \emph{spurious} CP close to the ones obtained from the expansion in HRG models, giving a CP value located at $(T,~\mu_B)\equiv (95,~815)$. 
Since the CEM has no first-order phase transition, this is again interpreted as the breakdown of the quadratic approximation at $\mu_B \sim 800$ MeV. 
Now that we have tested models describing the hadronic behavior, we will next look at models that describe the quark phase on the QCD phase diagram. 
\subsection{Nambu--Jona-Lasinio Model} \label{sec:njl}

We now analyze how our expansion performs in a quark-based model with degenerate bare quark masses. 
Specifically, we use the Nambu--Jona-Lasinio (NJL) model, the  Lagrangian of which reads:
\begin{equation}
    \mathcal{L}_{NJL} = \bar{\psi}(i\gamma^\mu \partial_\mu - m_0)\psi + G_S [(\bar{\psi}\psi)^2 + (\bar{\psi}i\gamma_5 \vec{\tau} \psi)^2] ,
\end{equation}
where $G_S$ is the scalar coupling, which gives the dynamical mass of the quark when coupled with the scalar condensate, and is also responsible for the pseudo-scalar coupling related to  pions. For simplicity, we do not include the effects of a vector coupling.

In a previous work, this model was employed to test whether the expansion of the entropy density contours up to second order could correctly locate the critical point~\cite{Marczenko:2025znt}. 
Here, we check if this expansion is valid by comparing the exact contours to those obtained via the quadratic extrapolation, both at positive and negative values of $\mu_B^2$.  

Using the mean field approximation, the effective thermodynamic potential is as follows:
\begin{equation}
    \Omega_{NJL} = \frac{(M-m_0)^2}{4G_S} + \Omega_{vac} + \Omega_{th},
\end{equation}
with
\begin{equation}
    \Omega_{vac} = \int^{\Lambda} \frac{d^3p}{(2\pi)^3} \epsilon_p, 
\end{equation}
where the integral for the vacuum part goes up to a cut-off parameter $\Lambda$. The thermal contribution to the thermodynamic potential is given by:
\begin{equation*}
    \Omega_{th} = 12T \int \frac{d^3 p}{(2\pi)^3} ( ln(1-f)+ ln(1-\bar{f})),
\end{equation*}
where $f$ and $\bar{f}$ are the Fermi-Dirac distributions for quarks and anti-quarks, and regularization is not required. 
We employ the model parameters from \cite{Buballa:2003qv}, namely $\Lambda = 0.5879$ GeV, $m_0 = 0.0056$ GeV, $G_s = 2.44/ \Lambda^2$
which yields a critical point at $(T,\mu_B) = (82, 966)$ MeV \cite{Buballa:2003qv}.

To obtain the pressure, one must first solve the gap equation to obtain the constituent quark mass \cite{Buballa:2003qv,Hatsuda:1994pi} 
\begin{equation*}
    \frac{\partial \Omega_{NJL}}{\partial M} = 0.
\end{equation*}
In our case, since the vector coupling is zero, this will be the only equation we need to solve. Once the gap equation is solved, the dynamical quark mass is used in the thermodynamic potential and the pressure at finite temperature and chemical potential is obtained as $ P(T,\mu) = -\Omega(T,\mu)$. From this pressure, one can obtain other thermodynamic quantities such as the entropy density and the second-order baryon susceptibility, which can be used to calculate the  quadratic coefficient $\alpha_2$ as well as the corresponding entropy-density contours.

When calculating the expansion coefficients in the model, we find no roots of the equation $\alpha_2''(T_0) = 0$, indicating that the extrapolated contours do not capture the phase transition. 
One likely reason is that the critical point in this model is relatively far from zero chemical potential, so a truncated quadratic approximation is not reliable at large baryon chemical potential. In fact, it was shown in \cite{Marczenko:2025znt} that the predictions from the extrapolated contours become more accurate when the CP is closer to $\mu_B=0$ in the NJL model. 

\begin{figure}[bth]
    \centering
    \includegraphics[width=0.9\linewidth]{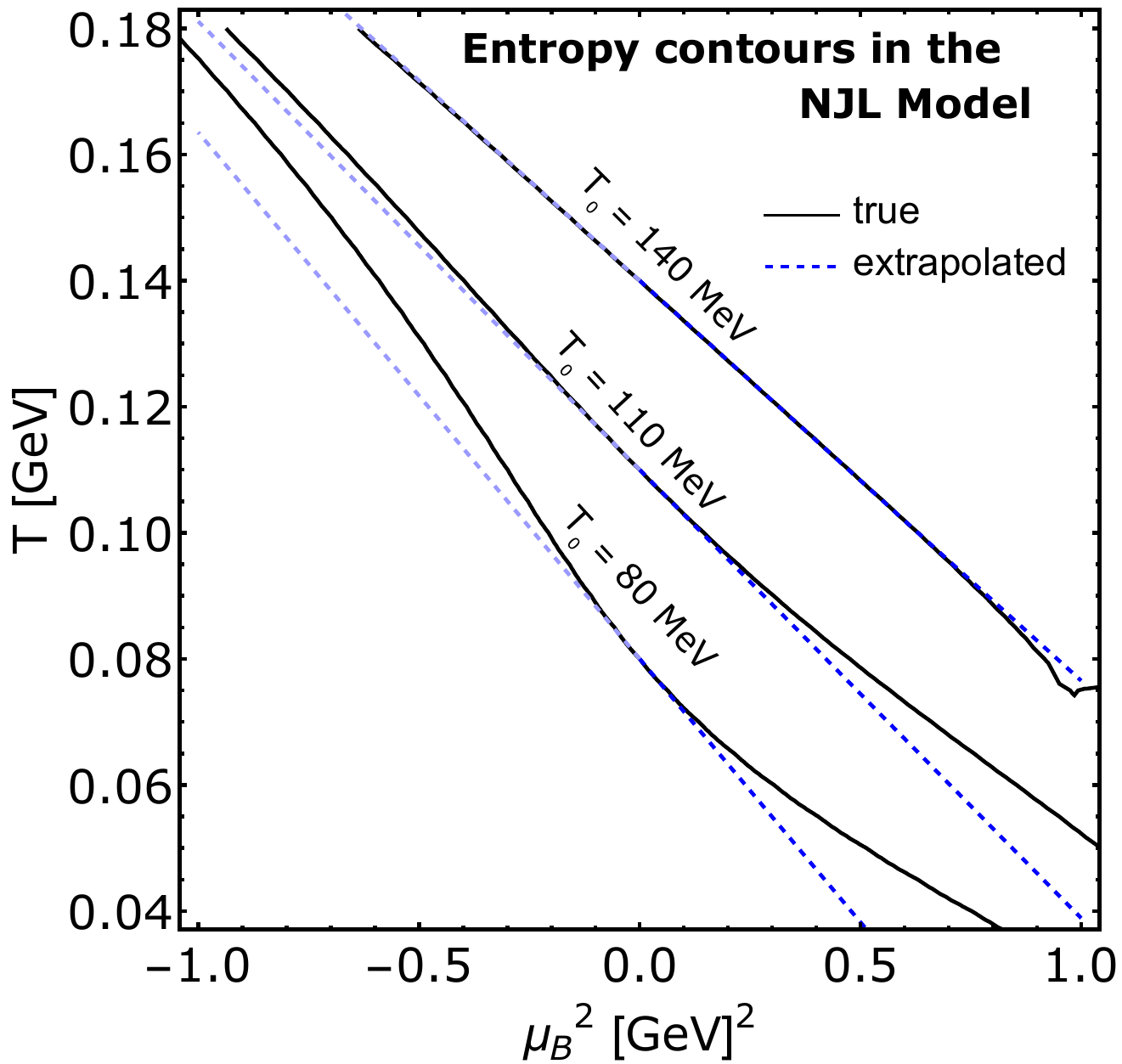}
    \caption{\justifying \small Full constant entropy density contours obtained from the NJL model (black, solid) compared to the truncated ones (blue, dashed) in the real and imaginary $\mu_B$ planes. }

    \label{fig:conts_NJL}
\end{figure}

This behavior can be understood by examining the contours of the model in the imaginary chemical potential plane, shown in Fig.~\ref{fig:conts_NJL}. The exact contours deviate from the approximate quadratic behavior, and the quadratic expansion fails to reproduce the true contours at large $|\mu_B|$. While contours starting from higher initial temperatures are reasonably well described at both real and imaginary $\mu_B$, the agreement between extrapolated and true contours becomes increasingly worse as $T_0$ decreases. This behavior, especially evident in the imaginary-$\mu_B$
 direction at low temperatures, shows that the quadratic expansion ceases to be valid for this model, as the constant- entropy contours emerging from low $T_0$ are not well approximated by a quadratic form in $\mu_B$.
\vspace{-1cm}

\begin{table*}[t]
\centering
\begin{tabular}{lccc}
\hline\hline
\textbf{Model} &
\textbf{Real CP $(T,\mu_B)$ [MeV]} & \hspace{0.2cm}
\textbf{Entropy/Enthalpy/Energy CP $(T,\mu_B)$ [MeV]} & \hspace{0.2cm}
\textbf{$T_{0,c}$ [MeV]} \\[1ex]
\hline
Ideal massless QGP &
--  &
-- / -- / -- &
-- \\

Ideal HRG &
--  &
$(86,817)$ / $(88,791)$ / $(86,777)$ &
$\sim 118$ \\

EV-HRG &
--  &
$(85,831)$ / $(87,805)$ / $(86,789)$ &
$\sim 115$ \\

vdW-HRG &
$(19,922)$ &
$(85,830)$ / $(87,805)$ / $(86,789)$ &
$\sim 115$ \\

CEM &
--  &
$(95,815)$ / $(96,802)$ / $(97,779)$ &
$\sim 111$ \\

BH Model &
$(103,599)$ &
$(104,637)$ / $(104,638)$ / $(104,611)$ &
$\sim 124$ \\

fRG--DSE QCD &
$(103,660)$ &
$(108,629)$ / $(107,624)$ / $(106,603)$ &
$\sim 133$ \\

NJL Model &
$(82,966)$ &
-- &
--\\

\textbf{Lattice QCD} &
\textbf{unknown} &
$(114,602)$ / $(112,606)$ / $(114,601)$ &
$\sim 141$ \\
\hline\hline
\end{tabular}
\caption{\justifying
Real and extrapolated critical point (CP) locations across models.
The contour-based CP column lists results obtained from
\textbf{constant entropy / constant enthalpy / constant energy density}
extrapolations, respectively.
The last column shows the critical $T_{0,c}$ for the constant entropy density
contour expansion.
Lattice values correspond to the expansion using the mean
parameterization of $s/T^3$ and $\chi_2^B$ given in Table~\ref{tab:mean_lattice}.
}
\label{tab:CP_model_compile}
\end{table*}

\vspace{1cm}
\subsection{Holographic Blackhole model} \label{sec:holography}
We now proceed to test our method on the Holographic EMD model, which contains a phase transition between two strongly coupled fluids \cite{Rougemont:2023gfz}. This section is based on the holographic gauge/gravity correspondence and its phenomenological application to QCD has been developed in Refs.~\cite{Gubser:2008yx,DeWolfe:2010he,DeWolfe:2011ts,Critelli:2017oub,Grefa:2021qvt}.%
\footnote{For independent work on similar models, see for instance \cite{Knaute:2017opk,Cai:2022omk,Li:2023mpv,Zhao:2023gur,Chen:2024ckb,Fu:2024nmw,Chen:2024epd,Jokela:2024xgz}.}
In this model,  the physical thermodynamic observables in a strongly coupled non-Abelian gauge theory (QCD) in flat four dimensions are obtained by solving the classical equations of motion of a higher-dimensional theory of gravity in asymptotically Anti-deSitter spacetime. The action in the EMD model is defined as:
\begin{align}
\label{eq:holoaction}
S = \frac{1}{2\kappa^2_{5}}\int_{\mathcal{M}_5} d^x \sqrt{-g}
\bigg(& R - \tfrac{(\partial_\mu \phi)^2}{2} - V(\phi) - \tfrac{f(\phi) F_{\mu\nu} F^{\mu\nu}}{4} \bigg)
\end{align}
where $\kappa_5^2 = 8\pi G_5$, while $G_5$ is the five-dimensional Newton’s constant.
 The real scalar field $\phi$ breaks the conformal invariance such as to mimic the running coupling of QCD, while $A_\mu$ is the Maxwell field that promotes global $U(1)_B$ symmetry associated with baryon-number conservation to a local symmetry in the bulk, and $F_{\mu \nu} = \partial_{\mu}A_{\nu}-\partial_{\nu}A_{\mu}$. The terms on the right hand side of Eq. \eqref{eq:holoaction}  correspond to the Einstein-Hilbert action, the kinetic and potential terms for the dilaton field and the Maxwell action, respectively. The dilaton-gauge coupling $f(\phi)$ couples the renormalization group flow to the baryonic current without breaking the $U(1)_B$ symmetry. 
\begin{figure}[bth]
    \centering
    \includegraphics[width=0.9\linewidth]{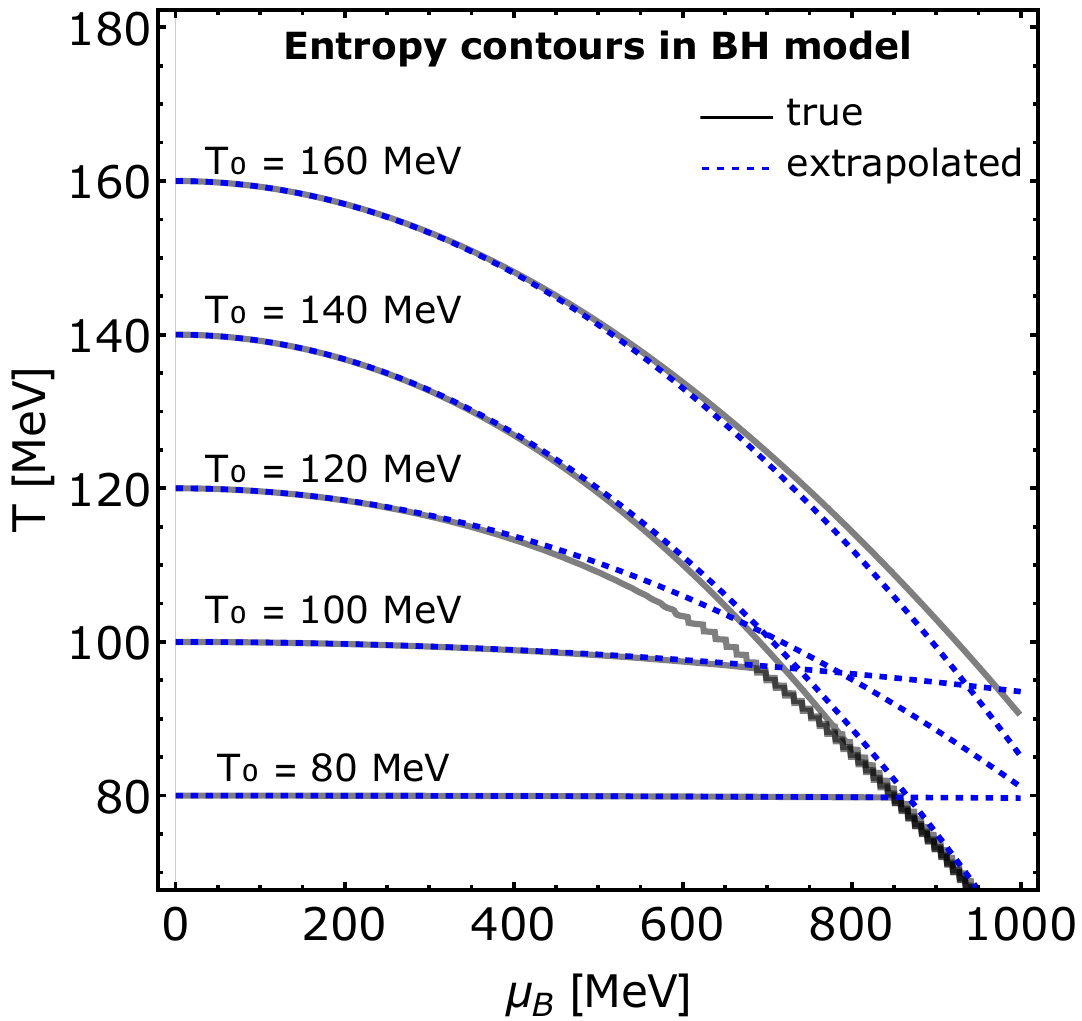}
    \caption{\justifying \small Full constant entropy density contours obtained from the Holographic Blackhole model (black, solid) compared to the extrapolated contours (blue, dashed) depicting a good match at quadratic order of the expansion.
    }
    \label{fig:conts_exp-holo}
\end{figure}

 Thermal states are described as black-hole solutions to the classical equations of motion, 
 each of them corresponding to a point in the $T-\mu_B$ phase diagram of QCD. 
 When mapped to QCD thermodynamics, the model gives a first-order phase transition between two strongly coupled fluids. One of the reasons to test the extrapolation in the Holographic Blackhole model is that it fixes the parameters in the functional form of $V(\phi)$ and $f(\phi)$ through lattice results for the entropy density $s$ and baryon susceptibility $\chi_2^B$, which are the necessary quantities needed in our extrapolation to follow the entropy density contours. 
 From these inputs, the EMD model is capable of predicting a variety of other lattice QCD results \cite{Critelli:2017oub,Grefa:2021qvt,Rougemont:2023gfz}. 
 Here we employ the so-called polynomial-hyperbolic Ansatz for $f(\phi)$ and $V(\phi)$ with best-fit parameters from Ref.~\cite{Hippert:2023bel}.

\begin{figure*}[bth]
    \centering
    \includegraphics[width=0.335\textwidth]{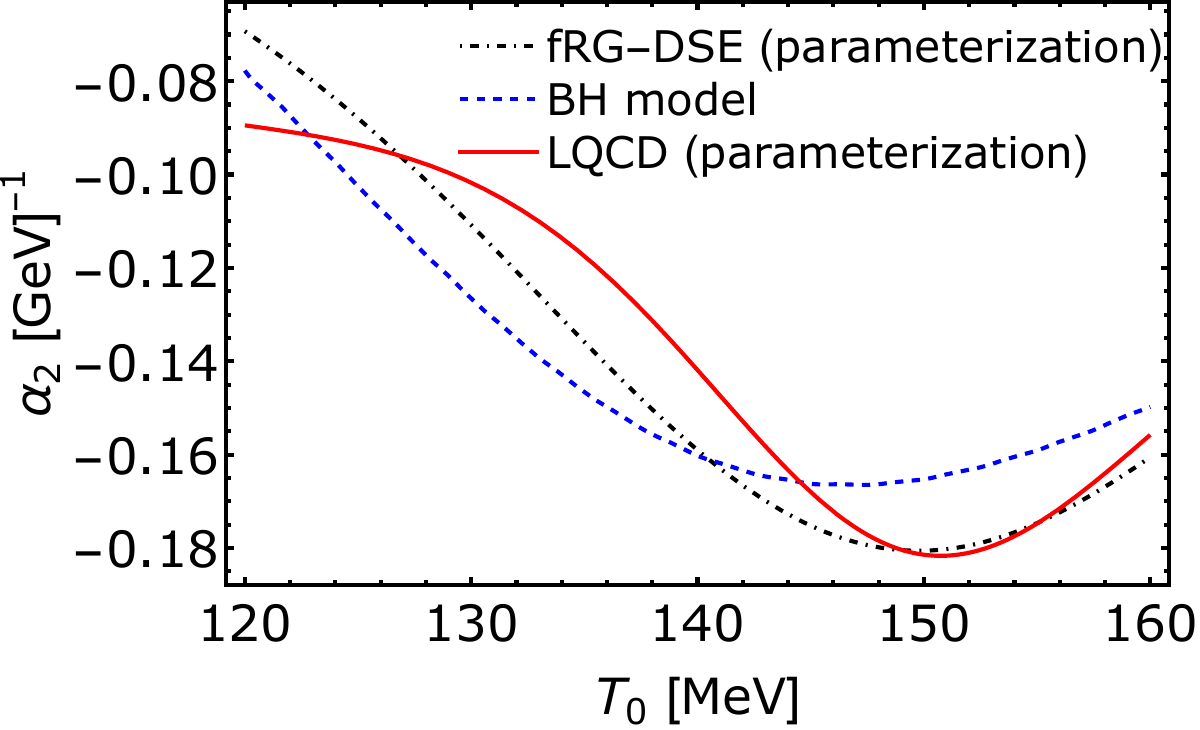}
    \includegraphics[width=0.32\linewidth]{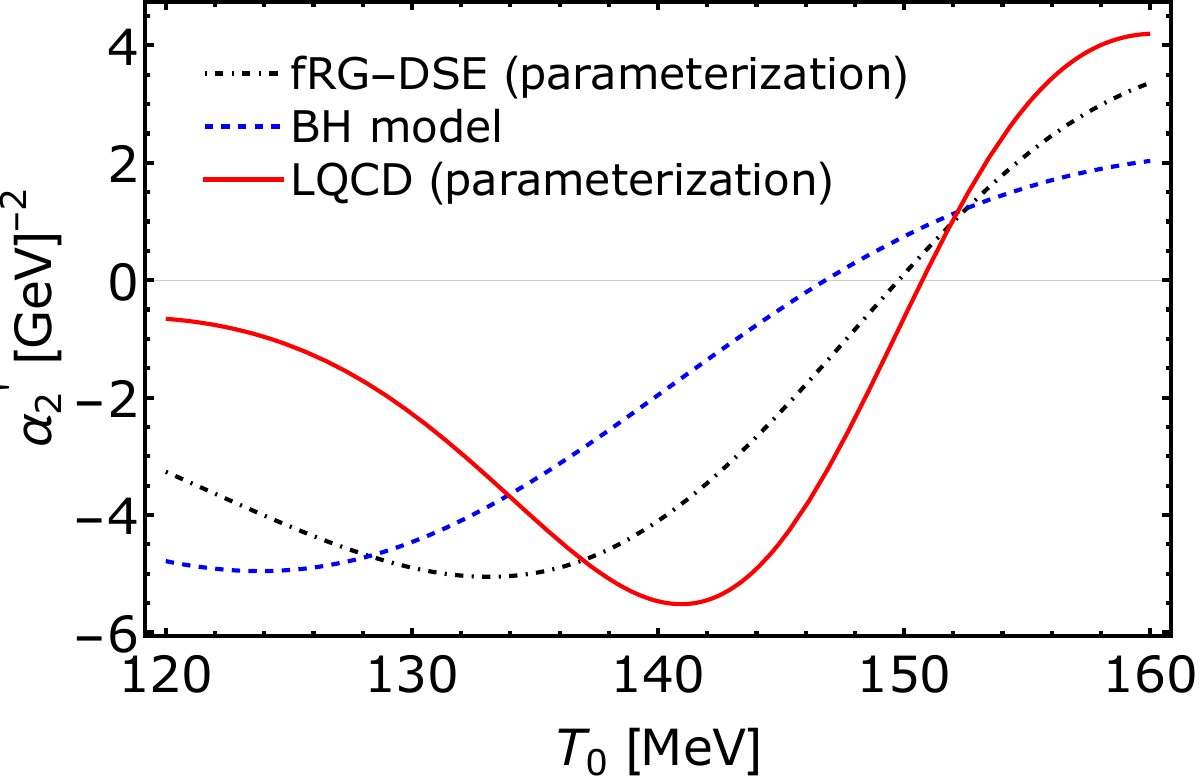}
    \includegraphics[width=0.33\linewidth]{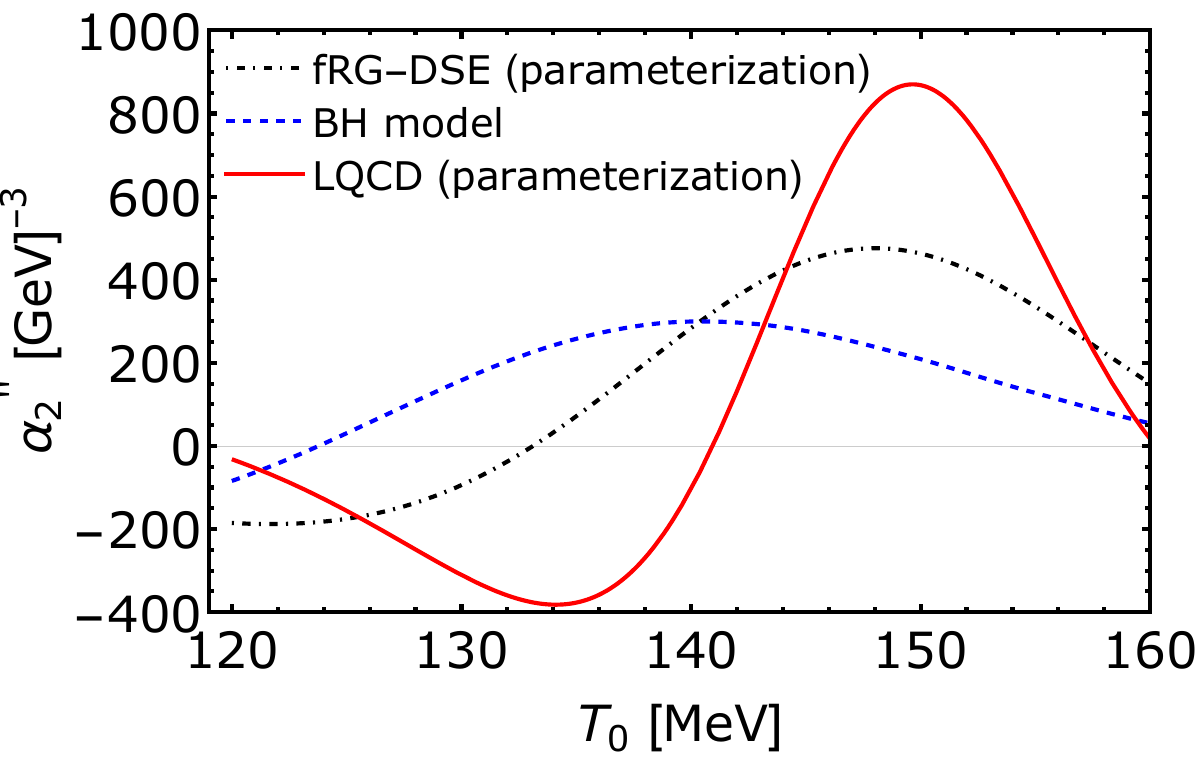}
    \caption{\justifying \small Expansion coefficient of constant entropy contours to $\mathcal{O}(\mu_B^2)$, along with its derivatives in different QCD-based models with a QCD phase transition, compared with the lattice QCD parameterization of the $\mu_B=0$ equation of state.
    }
\label{fig:alphas_lattice}
    \label{fig:conts_exp-holo1}
\end{figure*}

In Fig. \ref{fig:conts_exp-holo}, the reconstruction of constant entropy density contours from the expansion in Eq. (\ref{eq:TsO2}) is compared with the full contours from the holographic model. Even for an expansion that was truncated at 2nd order, the extrapolated contours show very good agreement with the actual contours of the model, indicating that the contours in the model itself are almost quadratic. We observe crossings of the contours suggesting a first-order phase transition in the model.
One observes crossings in the expanded contours, whereas no intersecting
contours appear in the Holographic Black Hole model. This is because the
Maxwell construction has already been performed on the equation of state
in the Holographic Black Hole model. As a result, one observes only the
first-order phase transition line and not the spinodal region, which is
what gives rise to the crossings in the constant entropy density contours.
If the Maxwell construction were not applied, one would indeed observe
crossings in the true contours of the model as well, and the expanded and
true contours of the model would overlap even beyond the critical point.
The critical point can be located by extrapolating entropy-density contours and using the temperature derivatives of $\alpha_2$, according to the criteria in  Eq.~\eqref{eq:spinodal} and Eq.~\eqref{eq:T0c}. The coefficient of expansion and its derivatives are shown in Fig.~\ref{fig:alphas_lattice}.
It is observed that $\alpha_2''$ vanishes at a certain value of $T_0$, suggesting the presence of a critical point, and that the corresponding chemical potential is real and finite. In this case, the critical point obtained by truncating the expansion at second order sits at $(T, \mu_B) = (104, 637)\,\text{MeV}$. The actual critical point in the model is located at $(T, \mu_B) = (103, 599)\,\text{MeV}$. Therefore, the results from the quadratic extrapolation are off by $1.4\%$ and $6.4\%$ in $T$ and $\mu_B$, respectively.  

The energy density contours yield a critical point at $(T, \mu_B) = (104, 611)\,\text{MeV}$, while the enthalpy density contours give $(T, \mu_B) = (104, 638)\,\text{MeV}$. Both sets of contours closely track the true critical region, further validating the extrapolation method and confirming that the temperature derivative of $\alpha_2$ provides a consistent and accurate signal of criticality across thermodynamic observables.
\subsection{Functional QCD} \label{sec:frg}
The continuum non-perturbative approaches of the functional renormalization group (fRG) and Dyson-Schwinger equations (DSE) can be combined to compute the equation of state (EoS) of QCD at finite temperature ($T$) and baryon chemical potential ($\mu_B$). In the fRG framework, one integrates quantum, thermal, and density fluctuations scale by scale through the Wetterich flow equation \cite{Wetterich:1992yh}, incorporating dynamical hadronization to account for emergent mesonic degrees of freedom. This approach enables consistent flows of couplings and propagators, while the DSE provides non-perturbative gap equations for quark and gluon fields. Combined, the fRG+DSE framework can be leveraged to yield an equation of state based on QCD, including pressure, entropy, and baryon density, across the crossover and first-order regions. 
This framework has been recently used to predict a critical point on the QCD phase diagram \cite{Fu:2022gou}, which is consistent with other effective approaches to QCD (see \cite{Fu:2019hdw, Gunkel:2021oya, Gao:2020fbl} for details). In particular, the recent study~\cite{Lu:2025cls} employs this combined framework to obtain an equation of state with a critical point at $(T,\mu_B)\rightarrow (103,660)$ MeV. 

Because the temperature derivatives of $s$ and $\chi_2^B$ required for applying the present method were not made  available in \cite{Lu:2025cls}, 
 we parameterized the entropy and $\chi_2^B$ for this dataset, using the same functional forms that were previously used for lattice data in \cite{Shah:2024img}, shown in Eqs. \eqref{eq:sT3param} and~\eqref{eq:chi2_para} in Section~\ref{sec:lattice} below.
 The expansionn coefficient using this parameterization is shown in Fig.~\ref{fig:alphas_lattice}, and the extrapolated CP is predicted at $(T,\mu_B)\rightarrow (108, 629)$ MeV.
Using enthalpy density and energy density contours, the extrapolated critical points are obtained at $(T,\mu_B)$=$(107, 624)$ MeV and $(106, 603)$ MeV, respectively. 
This should be compared to the actual critical point, found at $(T,\mu_B)= (103,660)$ MeV in Ref.~\cite{Lu:2025cls}. 
Due to $\alpha_2$ being extremely sensitive to the derivatives $s'$ and $\chi_2^{B'}$, one can obtain slightly different results based on the analysis performed on the data, and it becomes difficult to obtain the exact location for the extrapolated CP, unless these derivatives are calculated directly from the framework itself. Nonetheless, our analysis leads to an extrapolated critical $\mu_B$ within 10-15\% of the actual $\mu_{Bc}$ from Ref.~\cite{Lu:2025cls}.

The critical points obtained within different approaches, including fRG-DSE, and from various thermodynamic contours, are summarized in Table \ref{tab:CP_model_compile}. 
We stress that both the functional renormalization group and Holographic Blackhole models feature a critical point on the QCD phase diagram while successfully reproducing lattice-QCD results at vanishing chemical potential. Remarkably, the critical-point predictions from these two approaches fall within a similar range, and can be approximately reproduced by the expansion of constant-entropy contours truncated at the quadratic term -- especially in the holographic case, where the agreement is nearly exact. 
We further note that lattice QCD results at imaginary $\mu_B$ seem to confirm that the quadratic approximation indeed captures the behavior of the entropy density contours \cite{Borsanyi:2025dyp}.%
\footnote{It would  be interesting to study how the constant entropy density contours  at imaginary chemical potential in the Holographic Blackhole model and in the fRG–DSE approach, allowing for a more direct comparison with lattice QCD.}

\section{Thermodynamics from lattice QCD and applications}
\label{sec:lattice}

In the previous sections, we studied the expansion of constant entropy-density contours truncated at $\mathcal{O}(\mu_B^2)$ in various approaches.
Using different hadronic equations of state, we found that this expansion fails  to describe hadronic degrees of freedom when extrapolated to high $\mu_B$. 
Similarly to what was done in \cite{Marczenko:2025znt}, we have also examined the quark-based NJL model to find that the quadratic expansion of thermodynamic contours fails to capture the high-density phase transition of the model. Nonetheless, we notice that the NJL model does not accurately reproduce lattice QCD results at low density. 
We then turned to a Holographic Blackhole model \cite{Rougemont:2023gfz} and the fRG--DSE framework as more realistic descriptions of the deconfined and crossover regimes of QCD, able to reproduce lattice results at low $\mu_B$ while predicting a first-order phase transition at large baryon densities. 
Within these two approaches, we demonstrated that the expansion of constant-entropy-density contours truncated at $\mathcal{O}(\mu_B^2)$ provides an excellent description of thermodynamic observables in this region of the phase diagram.

Following these results, we now extend this methodology to lattice QCD in order to construct a full equation of state that incorporates a QCD critical point~\cite{Shah:2024img}. 
We reconstruct the pressure by integrating the entropy density in the temperature variable.
Since the expansion is not able to describe the hadronic physics correctly, we perform the integration from a non-zero temperature of $T = 80$~MeV, where we fix the integration constant by matching to a van der Waals HRG equation of state at a fixed temperature for each value of $\mu_B$. 
Doing so yields a well-defined thermodynamic equation of state in the low-temperature region. In this section, we describe in detail the methodology used to obtain the full thermodynamic equation of state from lattice QCD data within this framework. 
We then present the resulting thermodynamic quantities and discuss their behavior, particularly by looking at the critical point and the region of first-order phase transition.

\subsection{Contour expansion from lattice QCD}

In \cite{Shah:2024img}, we have already extracted the location of the QCD critical point from the quadratic expansion of  lattice QCD constant $s$ contours, taking into account the uncertainties in the lattice data.  
It was also shown that the entropy density extrapolated to finite $\mu_B$ was consistent with the already available lattice QCD equation of state for $\mu_B/T \leq 3.5$ \cite{Borsanyi:2021sxv}. When extended to higher values of $\mu_B$, the critical point location was found at $(T,\mu_B)$=$(114 \pm 7, 602 \pm 62)$ MeV at vanishing strange and electric charge chemical potentials ($\mu_S = \mu_Q = 0$).
 Here, we employ the expansion to obtain a lattice QCD-based equation of state with a critical point at finite $T$ and $\mu_B$. In addition to the entropy density studied previously, we compute the net-baryon density, pressure, speed of sound, and second-order baryon number susceptibility, and analyze their behavior in the vicinity of the critical point. Furthermore, we determine the Lee–Yang edge singularities from the expansion and investigate their behavior in the complex baryon chemical potential plane.

In order to apply the expansion to lattice data, the entropy density and the second-order baryonic susceptibility from the Wuppertal Budapest collaboration~\cite{Borsanyi:2013bia, Borsanyi:2021sxv} were parameterized to match the mean values of the dataset \cite{Shah:2024img}. 
A parameterization of these quantities is needed because their higher order temperature derivatives are required to obtain the expansion coefficient and its derivatives. 
The parameterization is given as follows:
\begin{align}
\label{eq:sT3param}
    \frac{s}{T^3} &= a \cdot \tanh \left( \frac{T - T_0}{d} \right) + b, \\
\chi_2^B(T) &= d_0 \left( \frac{2 \cdot m_p}{\pi \cdot x} \right)^{\frac{3}{2}} \cdot \frac{\exp\left(-\frac{m_p}{x}\right)}{1 + \left(\frac{x}{d_1}\right)^{d_2}} + d_3 \cdot \frac{\exp\left(-\frac{d_5^4}{x^4}\right)}{1 + \left(\frac{x}{d_1}\right)^{-d_2}}.
\label{eq:chi2_para}
\end{align}
Here $x = T /(200~\text{MeV})$ and $m_p = (938/200) \approx 4.7$ is
the proton mass in units of 200 MeV, as in Ref.~\cite{Kahangirwe:2024cny}.
Propagating uncertainties of the lattice QCD input through the parametrization,
the critical point location was found to be $(T,\mu_B)$=$(114 \pm 7, 602 \pm 62)$ MeV \cite{Shah:2024img}. The mean value for the critical point was obtained through the mean value of the parameters (given in Table \ref{tab:mean_lattice}), such that the resulting lines pass through the lattice QCD data points. 
In Ref.~\cite{Shah:2024img}, it was also shown that using smoothing splines instead of this parameterization yields consistent results within errors, indicating that the particular parametric form we used does not bias the results appreciably.
\begin{table}[]
    \centering
  \begin{tabular}{|c|c|}
    \hline
        Parameter & Value \\
        \hline
        $a$ & 5.65608 \\ 
        $b$ & 6.43026 \\ 
        $T_0$ & 163.681 \\
        $d$ & 43.3516 \\
        \hline
    \end{tabular}
    \hspace{1cm}
     \begin{tabular}{|c|c|}
    \hline
        Parameter & Value \\
        \hline
        $d_0$ & 3.60763 \\ 
        $d_1$ & 0.750215 \\ 
        $d_2$ & 21.1553 \\
        $d_3$ & 0.330518 \\
        $d_5$ & 0.758584 \\
        \hline
    \end{tabular}
    \caption{\justifying Mean parameter values for $s/T^3$ (left) and $\chi_2^B$ (right) describing lattice QCD data using the parameterization given in Eq.\eqref{eq:sT3param} and Eq.\eqref{eq:chi2_para} respectively from Ref.~\cite{Shah:2024img}.
    The covariance matrices are available in Ref.~\cite{Shah:2024img}.
    }
    \label{tab:mean_lattice}
\end{table}

As a consistency check, before constructing the equation of state from constant entropy-density contours, we first examine the location of the critical point (CP) obtained from constant enthalpy-density and constant energy-density contours. In an exact equation of state, the CP should be independent of the choice of the thermodynamic variable used to define the contours. However, since the expansion is truncated at second order in $\mu_B$, deviations in the extracted CP location may arise when different thermodynamic quantities are employed. Using Eq.~\eqref{eq:alpha_2_energy}, we obtain a CP using linear error propagation of lattice uncertainties resulting the location at $(T,\mu_B) \rightarrow (113 \pm 7,584 \pm 61)$ MeV from the energy-density contours, while employing the enthalpy-density expansion in Eq.~\eqref{eq:alpha_2_enthalpy} yields $(T,\mu_B) \rightarrow (113 \pm 6,606 \pm 58)$ MeV. Despite the second-order truncation, the CP locations extracted from these contours are in close agreement and lie within the 68\% confidence interval of the CP determined from constant–entropy-density contours~\cite{Shah:2024img}.

\subsection{Equation of State}

Hydrodynamic simulations play a central role in the theoretical interpretation of the medium created in heavy-ion collisions, as they model the space-time evolution of the strongly interacting matter produced in these experiments~\cite{Luzum:2008cw,Carzon:2023zfp,MUSES:2023hyz,Dore:2020jye,Stephanov:2017ghc,Jeon:2015dfa,Karpenko:2015xea,Gale:2012rq}. 
In particular, complete simulations are required to realistically describe the effects of the QCD critical point on the evolution of the system. 
While standard relativistic viscous hydrodynamics requires important modifications near the QCD critical point, where fluctuations and non-trivial transport phenomena become significant, effects of criticality should also enter these simulations via the equation of state (EoS),  which should incorporate all available theoretical insights.
Here, we use our expansion along contours of constant entropy density to construct a lattice-based equation of state presenting a critical point at high baryon density.

Before obtaining the thermodynamics from the entropy density contours through lattice data, it is important to recall the type of structure we might expect when calculating the equation of state. If one could perform first principle calculations of the QCD equation of state in the thermodynamic limit, in the first order phase transition region one would see the most stable solution, namely a discontinuity in the entropy density. In the case of our expansion, as it is analytic and truncated at second order in $\mu_B$, one observes the metastable and unstable branches in the equation of state, reflecting a mean-field-type behavior. Thus, the entropy is expected to have a multi-valued behavior, and the other thermodynamic quantities as well.
\begin{figure*}
    \centering

    \includegraphics[width=\linewidth]{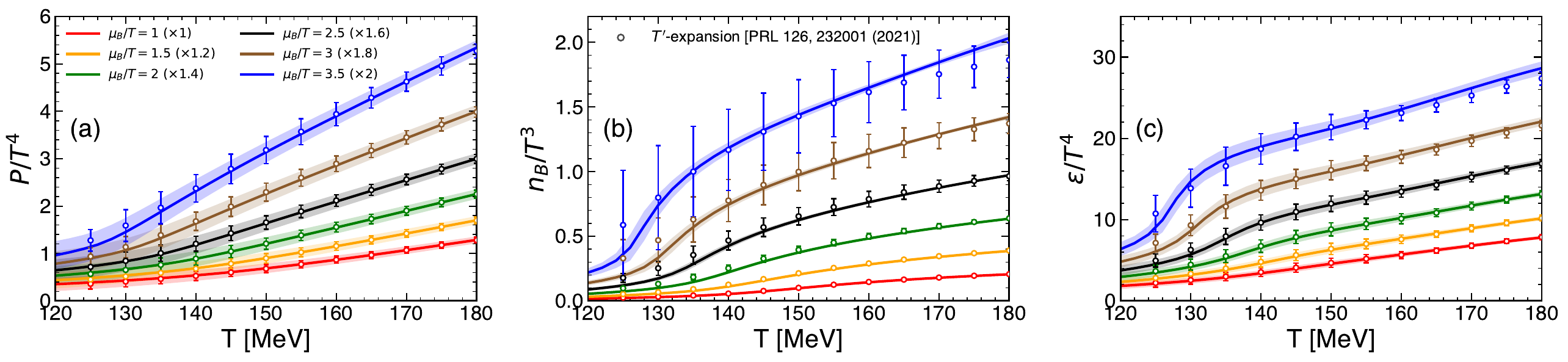}

    \caption{\justifying Scaled (a) pressure, (b) net baryon density and (c) energy density as functions of the temperature at constant values of $\mu_B/T$ obtained from our analysis (solid lines) compared with the $T'-$ expansion scheme results (error bars) from Ref.~\cite{Borsanyi:2021sxv}.
    The bands depict the error propagation of the lattice QCD input data.}
    The $\mu_B/T$ slices are scaled by numerical factors to avoid overplotting.
    \label{fig:muT_compare}
\end{figure*}

Applying the parameterization from the previous section and the expansion Eq. \ref{eq:TsO2}, one can calculate the entropy density in the $(T,\mu_B)$ plane. 
At this order, each contour is labeled by its intercept $T_0$ at $\mu_B = 0$, and for a given point $(T,\mu_B)$ one obtains $T_0(T,\mu_B)$ by solving the implicit relation $T = T_s(\mu_B;T_0)$ in Eq.~\eqref{eq:TsO2}.
However, the entropy as a function of $T$ and $\mu_B$ itself does not contain full thermodynamic information about the system.
Instead, it is the pressure, which has $T$ and $\mu_B$ as its natural variables.
Entropy and pressure are related through a thermodynamic identity $s(T,\mu_B) = [\partial P(T,\mu_B) / \partial T]_{\mu_B}$. Therefore, the pressure can be reconstructed by 
integrating the entropy density with respect to the temperature:
\begin{equation}
\label{eq:P1}
    \int_{T_{\rm low}}^T s(T',\mu_B) dT' = P(T,\mu_B) - P(T_{\rm low}, \mu_B),
\end{equation}
where $T_{\rm low}$ is some temperature below $T$.
To fix the integration constant $P(T_{\rm low}, \mu_B)$ we set $T_{\rm low} = 80$ MeV and use the QvdW-HRG model to calculate $P(T_{\rm low}, \mu_B)$.

The integral in Eq.~\eqref{eq:P1} can be simplified.
We note that $s(T,\mu_B) = s_0[T_0(T,\mu_B)]$ where $T_0(T',\mu_B)$ is determined through Eq.~\eqref{eq:TsO2} (where $T \equiv T_s$) in implicit form.
Making a variable change $T' \to T_0'$ as using, based on Eq.~\eqref{eq:TsO2}, that
\begin{align}
    \left(\frac{dT}{dT_0}\right)_{\mu_B} = 1 + \frac{\alpha_2'(T_0) \mu_B^2}{2},
    \label{eq:dT0}
\end{align}

we can write the pressure as
\begin{align}
\label{eq:P2}
    P(T,\mu_B) & = P(T_{\rm low}, \mu_B) + \int_{T_{0}^{\rm low}}^{T_{0}^{\rm up}} s_0(T_0') dT_0'  \nonumber \\
    & \quad + \int_{T_{0}^{\rm low}}^{T_{0}^{\rm up}} s_0(T_0') \frac{\alpha_2'(T_0) \mu_B^2}{2} dT_0'.
\end{align}
Here $T_{0}^{\rm low} \equiv T_{0}(T_{\rm low},\mu_B)$ and $T_{0}^{\rm up} \equiv T_{0}(T,\mu_B)$.

The first integral in Eq.~\eqref{eq:P2} yields the pressure difference at $\mu_B = 0$, i.e. 
\begin{align}
\int_{T_{0}^{\rm low}}^{T_{0}^{\rm up}} s_0(T_0') dT_0' = p_0(T_{0}^{\rm up}) - p_0(T_{0}^{\rm low})
\equiv \left. p_0(T_0) \right|_{T_{0}^{\rm low}}^{T_{0}^{\rm up}}.
\end{align}
To evaluate the second integral, we apply integration by parts.
One obtains
\begin{align}
\label{eq:P3}
    & P(T,\mu_B)  = P(T_{\rm low}, \mu_B) + \left. p_0(T_0) \right|_{T_{0}^{\rm low}}^{T_{0}^{\rm up}}  \nonumber \\
    & \quad + \frac{\mu_B^2}{2} \left. \left[s_0 \alpha_2 \right] \right|_{T_{0}^{\rm low}}^{T_{0}^{\rm up}}
    - \frac{\mu_B^2}{2} \int_{T_{0}^{\rm low}}^{T_{0}^{\rm up}} s_0'(T_0') \alpha_2(T_0') dT_0'~.
\end{align}

To evaluate the remaining integral in Eq.~\eqref{eq:P3} we use the fact that $\alpha_2(T_0) = - [T_0^2 \chi_2^B(T_0)]' / s_0'(T_0)$.
This leads to the final expression for the pressure which avoids numerical integration
\begin{align}
\label{eq:Pfinal}
    P(T,\mu_B) & = P(T_{\rm low}, \mu_B) + \left. p_0(T_0) \right|_{T_{0}^{\rm low}}^{T_{0}^{\rm up}}  \nonumber \\
    & \quad + \frac{\mu_B^2}{2} \left. \left[s_0(T_0) \alpha_2(T_0) + T_0^2 \chi_2^B(T_0)\right] \right|_{T_{0}^{\rm low}}^{T_{0}^{\rm up}}.
\end{align}

The baryon density is obtained from the thermodynamic relation $ n_B (T,\mu_B) = \left[\frac{\partial P(T,\mu_B)}{\partial\mu_B}\right]_T$.
Applying this derivative to Eq.~\eqref{eq:Pfinal} one obtains:
\begin{align}
\label{eq:nB1}
& n_B(T,\mu_B) = n_B(T_{\rm low},\mu_B) + \left. \left[ s_0(T_0) \left( \frac{\partial T_0}{\partial \mu_B} \right)_T \right] \right|_{T_{0}^{\rm low}}^{T_{0}^{\rm up}}\nonumber \\
& \quad + \mu_B \left. \left[s_0(T_0) \alpha_2(T_0) + T_0^2 \chi_2^B(T_0)\right] \right|_{T_{0}^{\rm low}}^{T_{0}^{\rm up}} \nonumber \\
& \quad + \frac{\mu_B^2}{2} \left. \left[s_0(T_0) \alpha_2'(T_0) \left( \frac{\partial T_0}{\partial \mu_B} \right)_T\right] \right|_{T_{0}^{\rm low}}^{T_{0}^{\rm up}}.
\end{align}
To calculate the derivative $\left(\frac{\partial T_0}{\partial \mu_B}\right)_T$, we differentiate the implicit relation $T= T_0 + \alpha_2(T_0) \mu_B^2/2$ at fixed $T$:
\begin{equation}
\begin{gathered}
0=
\left( \frac{\partial T_0}{\partial \mu_B} \right)_T
+\mu_B\,\alpha_2(T_0)
+\frac{\mu_B^2}{2}\,\alpha_2'(T_0)
\left( \frac{\partial T_0}{\partial \mu_B} \right)_T,\\[0.4em]
\left( \frac{\partial T_0}{\partial \mu_B} \right)_T
=
-\frac{\mu_B\,\alpha_2(T_0)}
{1+\frac{\mu_B^2}{2}\alpha_2'(T_0)} 
\end{gathered}
\end{equation}
Plugging this expression into Eq.~\eqref{eq:nB1}, after simplifications, we obtain a compact final expression for $n_B(T,\mu_B)$:
\begin{align}
\label{eq:nB}
n_B(T,\mu_B) = n_B(T_{\rm low},\mu_B) + \mu_B \left. \left[T_0^2 \chi_2^B(T_0)\right] \right|_{T_{0}^{\rm low}}^{T_{0}^{\rm up}}.
\end{align}
Using the above equation, one obtains the baryon density as a function of temperature and baryon chemical potential.
For completeness, we also write here the $\mu_B$ derivative of the entropy density $s(T,\mu_B)$:
\begin{align}
    \left (\frac{\partial s}{\partial \mu_B} \right)_{T} = - \mu_B \frac{\alpha_2 (T_0) s'(T_0)}{1 + \frac{\alpha_2'(T_0) \mu_B^2}{2}}.
    \label{eq:dsdmuB_final}
\end{align}

From the baryon density, pressure and  entropy density, one can calculate the energy density, which is given by Eq. \eqref{eq:edens}. Our final results for the normalized pressure, energy density and baryon density are shown as functions of temperature at constant values of $\mu_B/T$ in Fig. \ref{fig:muT_compare}. The solid lines with bands are the values obtained in this analysis, while the points with error bars are lattice QCD-based results from Ref. \cite{Borsanyi:2021sxv} based on the $T'$ expansion scheme. 
We observe that both expansion schemes are in agreement up to $\mu_B/T \leq 3.5$. 

\begin{figure*}
    \centering

    \includegraphics[width=\linewidth]{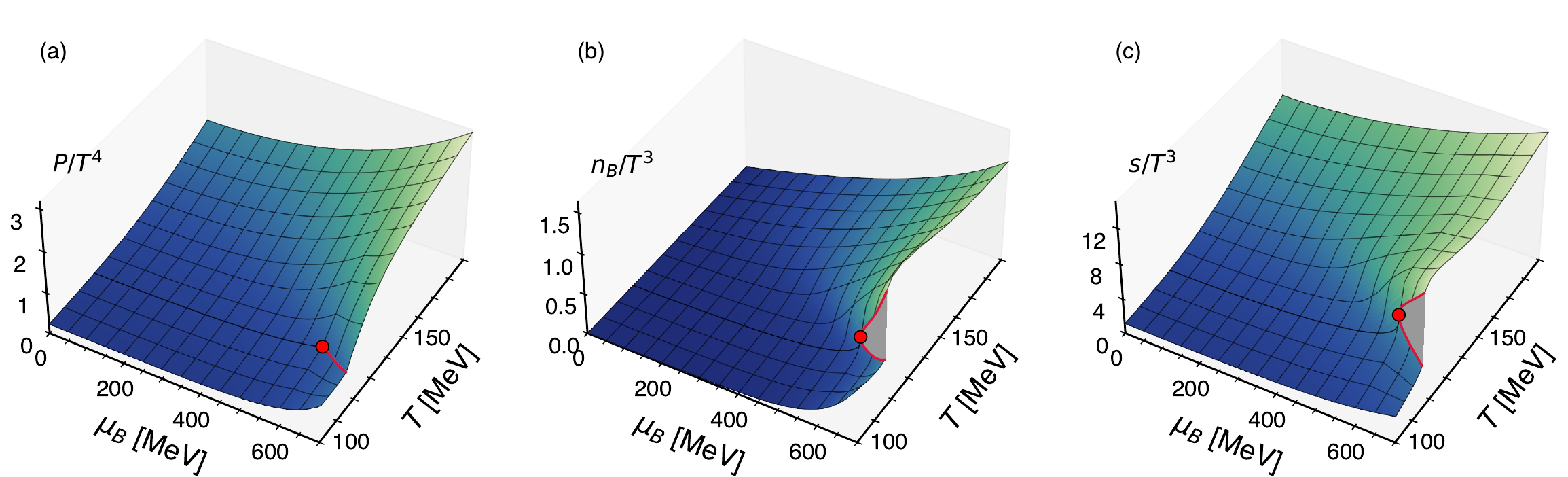}
    \caption{\justifying (a) Normalized pressure $P/T^4$, (b) net baryon density $n_B/T^3$,
    and (c) entropy density $s/T^3$ as functions of temperature and baryon
    chemical potential.
    For $\mu_B > \mu_{B,c}$ the surfaces correspond to the thermodynamically stable
    branches selected by the Maxwell construction.
    The solid red lines depict the boundaries of the first-order phase transition
    along the coexistence line.
    The red circle marks the critical point, where the first-order transition ends
    and the discontinuities in $n_B$ and $s$ close.}
    \label{fig:3D_EoS}
\end{figure*}

\begin{figure*}
    \centering

    \includegraphics[width=\linewidth]{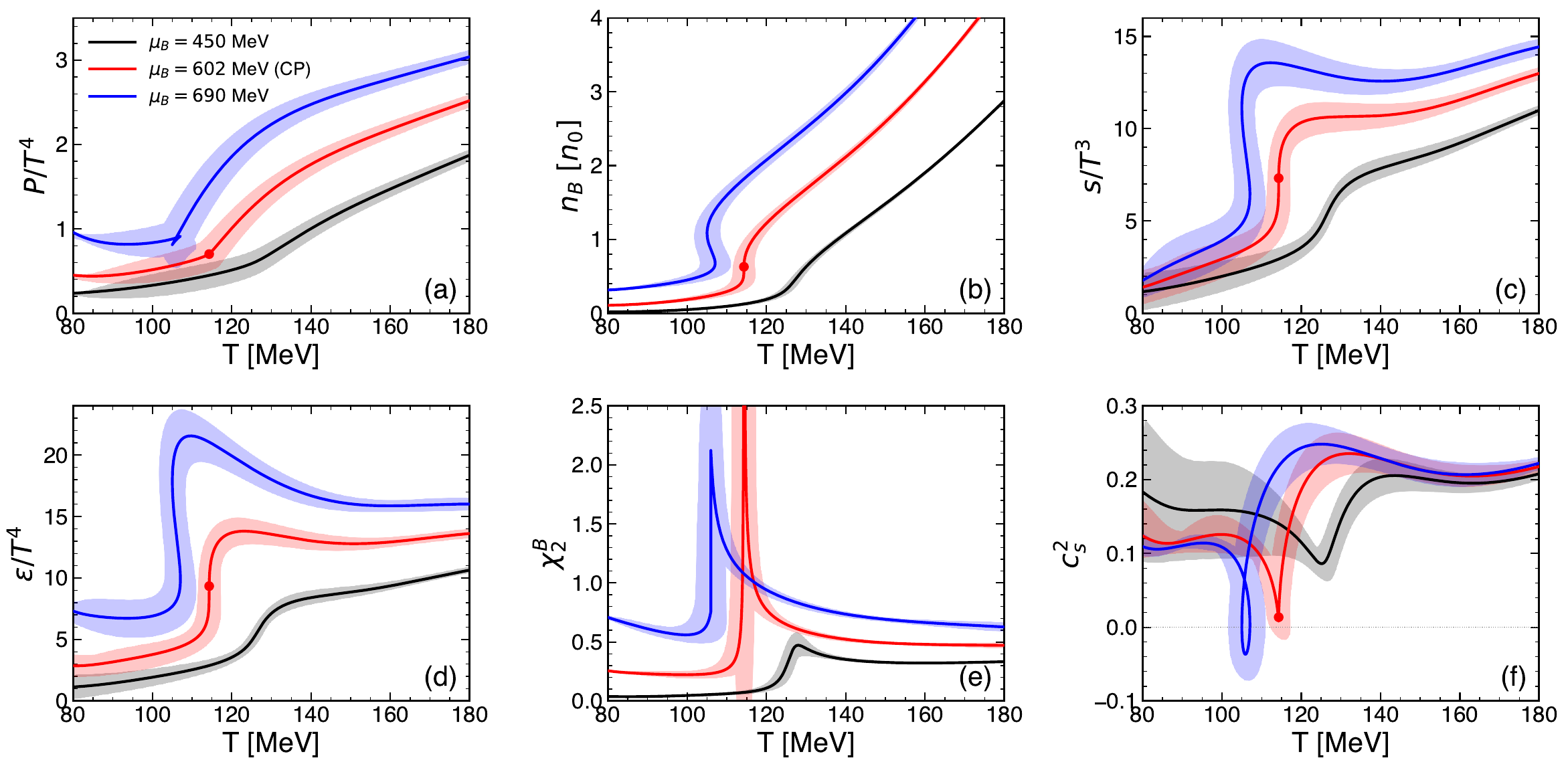}

    \caption{\justifying Equation of state properties obtained in our analysis. The different quantities are shown as functions of the temperature for different constant values of the baryon chemical potential. 
    The quantities are (a) the scaled pressure, (b) baryon number density (in units of nuclear saturation density),  (c) scaled entropy density, (d) scaled energy density, (e) second order baryon number susceptibility (bottom, middle) and (f) square of the speed of sound, in the crossover ($\mu_B = 450$~MeV), critical point ($\mu_B = 602$ MeV) and first order phase transition ($\mu_B = 690$ MeV)  regions.
    The baryon susceptibility plot has Maxwell construction applied for the $\mu_B = 690$ MeV, the other plots are obtained without the Maxwell construction.
    The bands depict the error propagation of the lattice QCD input data.
    }
    \label{fig:EoS}
\end{figure*}

To further illustrate the resulting equation of state in the $(T,\mu_B)$ plane, Fig.~\ref{fig:3D_EoS} shows the normalized pressure, baryon density, and entropy density surfaces as functions of temperature and baryon chemical potential. In the region of the first-order phase transition, the Maxwell construction is applied and only the thermodynamically stable solution is displayed. 
Finally, Fig.~\ref{fig:EoS} shows the behavior of different thermodynamic quantities as functions of temperature at different values of the baryon chemical potential.
For $\mu_B > \mu_{Bc}$, the multiple solutions associated with the first-order phase transition are kept intact.


The above procedure allows us to analyze the structure of the CP and the associated mixed phase region in the vicinity of the CP in the $T$-$n_B$ plane, in addition to the $T$-$\mu_B$ plane studied before.
We note that baryon density at the CP is $n_{B,c} \approx 0.6 \,n_0$, where $n_0 = 0.16$~fm$^{-3}$ is the normal nuclear density.
At subcritical temperatures, $T \lesssim T_c$, the first-order phase transition leads to the emergence of a mixed-phase region in the $T$-$n_B$ plane.
As the temperature is decreased, the baryon density of both the low- and high-density phases \emph{increase} relative to $n_{B,c}$.
This feature is markedly different from the inverted U-shape commonly observed in mean-field models without explicit hadronic degrees of freedom, such as NJL, but consistent with the ``banana-like" shape of the coexistence region modeled in Ref.~\cite{Steinheimer:2013gla}.
The expected shape of the coexistence region was further discussed in~\cite{Steinheimer:2013xxa}, where the liquid-gas type coexistence region in mean-field models was criticized as unphysical, leading, in particular, to the coexistence of vacuum and quark droplets when extended to $T = 0$.
For a CP located at a baryon density smaller than the normal nuclear density $n_0$, as is the case in our study, the shape of the coexistence region thus necessarily has to be different from the inverted U-shape, and this is what we observe here.

Using Eq. \eqref{eq:nB}, the second order baryon number susceptibility can be obtained by taking the derivative of baryon density with respect to $\mu_B$, namely $T^2 \chi_2^B(T,\mu_B) = (\partial n_B / \partial \mu_B)_T$. One obtains
\begin{align}
T^2\chi_2^B (T,\mu_B) 
& = T_{\rm low} ^2\chi_2^B (T_{\rm low},\mu_B) + \left. \left[T_0^2 \chi_2^B(T_0)\right] \right|_{T_{0}^{\rm low}}^{T_{0}^{\rm up}} \nonumber \\
& + \mu_B^2 \left. \left[\frac{s_0'(T_0) [\alpha_2(T_0)]^2}{1 + \frac{\mu_B^2}{2} \alpha_2'(T_0)}\right] \right|_{T_{0}^{\rm low}}^{T_{0}^{\rm up}}.
\label{eq:chi2B2}
\end{align}
Using the above equation, $\chi_2^B$ as a function of the temperature at constant $\mu_B$ is obtained, and shown in Fig. \ref{fig:EoS}. 
The resulting $\chi_2^B$ diverges at the critical point, as expected. Thereafter, in the first-order phase transition, we apply Maxwell construction 
to obtain and depict in Fig.~\ref{fig:EoS} only the stable branch solution for this quantity. 
The adiabatic squared speed of sound  reads~\cite{Floerchinger:2015efa}:
\begin{equation}
    c_s^2(T,\mu_B) = \left( \frac{\partial p}{\partial \epsilon} \right)_{s/n} = \frac{n^2 \frac{\partial^2p}{\partial T^2} - 2sn \frac{\partial^2p}{\partial\mu_B \partial T} + s^2 \chi_2^B}{(\epsilon+p)\left( \frac{\partial^2 p}{\partial T^2}\chi_2^B - \left( \frac{\partial^2 p}{\partial \mu_B \partial T}\right)^2   \right)}.
\end{equation}
Notably, $c_s^2$ stays finite at the CP in our model.
This reflects the mean-field universality class that our construction provides, and different from the expected 3D-Ising universality class behavior, where the value of the speed of sound would be exactly zero at the CP \cite{folk1999critical,onuki2002phase}. 
Moving beyond the critical chemical potential, we enter the first-order phase transition region, where both metastable and unstable branches of the speed of sound appear. 
In particular, the formal single-phase expression for $c_s^2$ yields negative values within the spinodal region, reflecting the mechanical instability.  

\begin{table}[t]
\centering
\begin{tabular}{|c|c|}
\hline
\textbf{Quantity} & \textbf{Value at the CP} \\
\hline
$T_0$ [MeV]                   & $140.94 \pm 1.95$\\
\hline
$T$ [MeV]                   & $114.29 \pm 6.85$\\
\hline
$\mu_B$ [MeV]               & $602.06 \pm 62.05$\\
\hline
$P/T^4$             & $0.70 \pm 0.19$ \\
\hline
$s/T^3$             & $6.95 \pm 0.82$ \\
\hline
$n_B/n_0$            & $0.59 \pm 0.13$\\
\hline
$s/n_B$            & $14.36 \pm 4.25$\\
\hline
$\epsilon/T^4$ & $8.67 \pm 1.79$\\
\hline
\hline
\end{tabular}
\caption{\justifying Thermodynamic quantities evaluated at the critical point (CP).
The displayed errors propagate the uncertainties in the lattice QCD input.
}
\label{tab:CP_thermo_values}
\end{table}

\subsection{ Isentropic Trajectories}

\begin{figure}[bth]
    \centering
    \includegraphics[width=0.76\linewidth]{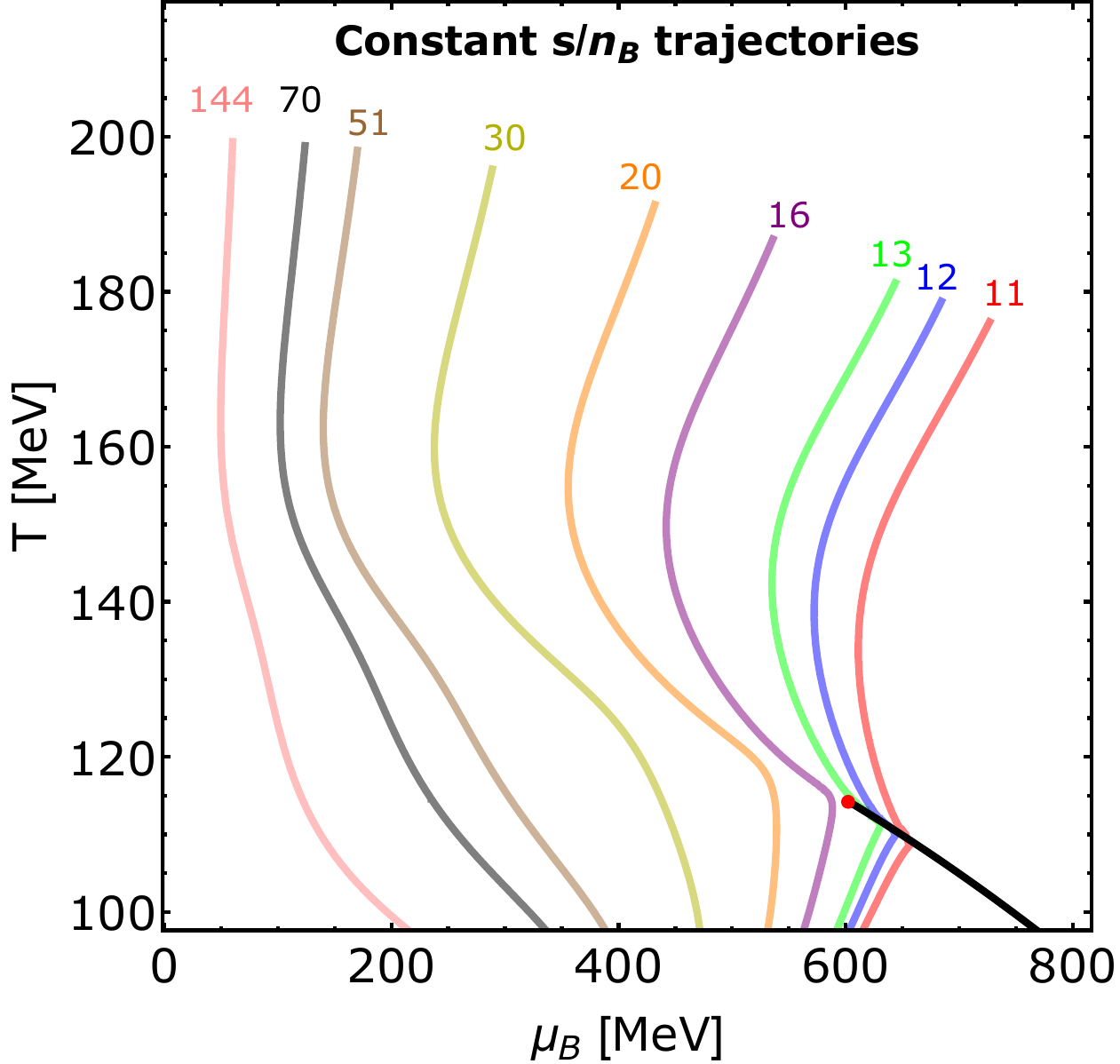}
    \caption{\justifying Several isentropic trajectories in the $(T,\mu_B)$ plane converging towards the critical point and line of first order phase transition.
    } 
    \label{fig:sonB}
\end{figure}

Utilizing the entropy density and net-baryon density obtained in our analysis, we can calculate the isentropic trajectories of constant $s/n_B$ across the phase diagram. 
They are shown in Fig. \ref{fig:sonB} for a few different values of $s/n_B$. 
At small $\mu_B$ the trajectories in our constructed EoS are similar to various lattice-based calculations in the literature~\cite{Gunther:2017sxn,Vovchenko:2017gkg,Grefa:2021qvt}.
At higher densities we observe a focusing behavior due to the critical point, similar to the one discussed recently in \cite{Pradeep:2024cca}.
When traversing a first-order coexistence line, the trajectories exhibit a jump before exiting in the hadronic phase.
This is consistent with expectations for the QCD phase transition and in contrast to a liquid-gas type transition, which has locally incoming isentropes only~\cite{Wunderlich:2016aed}.

The isentropic trajectories are relevant for heavy-ion collisions, as they approximate the trajectories that the system created in these collisions would follow during its evolution when viscous effects can be neglected. 
Hence, these trajectories can provide useful information on the effect of the CP in heavy-ion experiments. 
However, it is important to note that the system studied here corresponds to $\mu_S = 0$ and $\mu_Q = 0$, while the system created in heavy ion collisions has global strangeness neutrality and $n_Q/n_B \approx 0.4$. 
This difference is insignificant at small chemical potentials but may become sizable at $\mu_B \gtrsim 600$ MeV~\cite{Fu:2018qsk,Lysenko:2024hqp}.

\vspace{-1cm}
\vspace{0.5cm}
\subsection{Lee-Yang Edge Singularities}
\begin{figure*}[bth]
    \centering
    \includegraphics[width=0.4\linewidth]{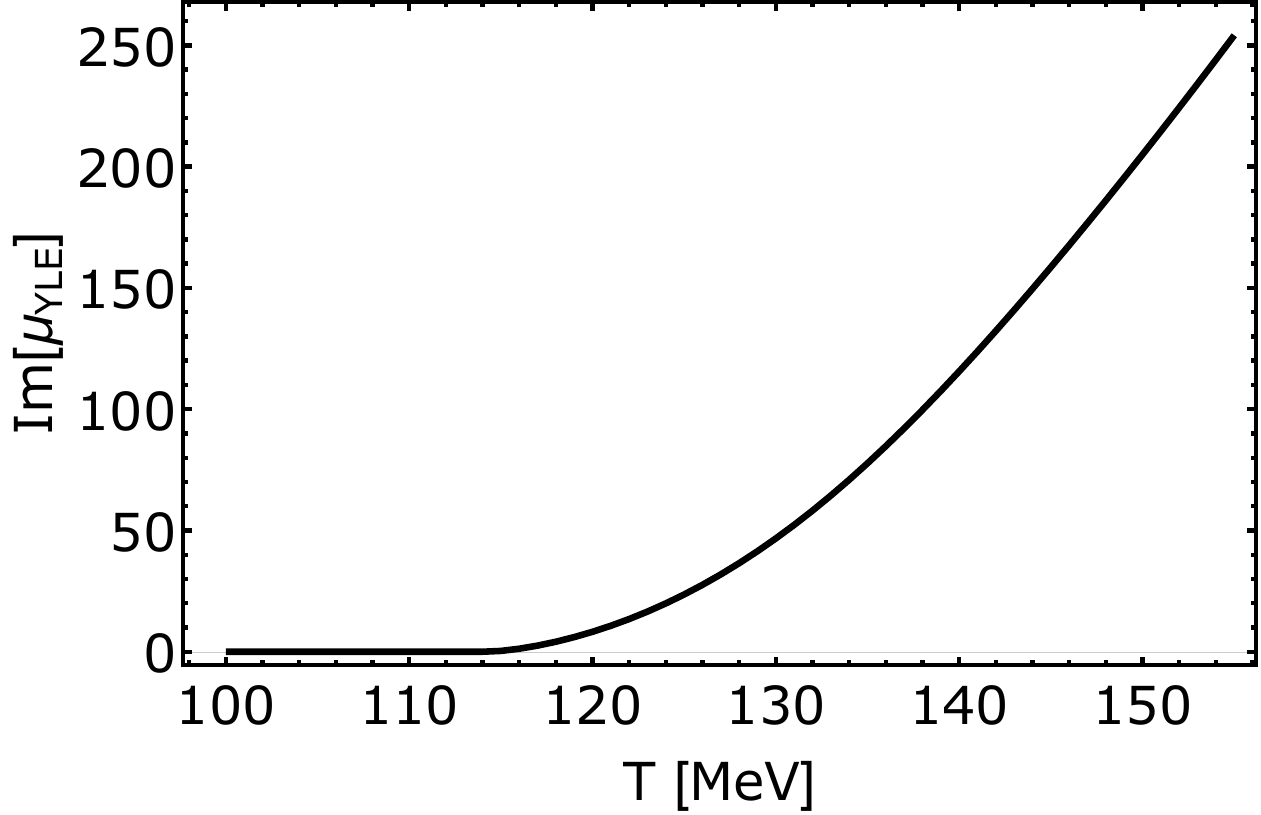}
    \includegraphics[width=0.4\linewidth]{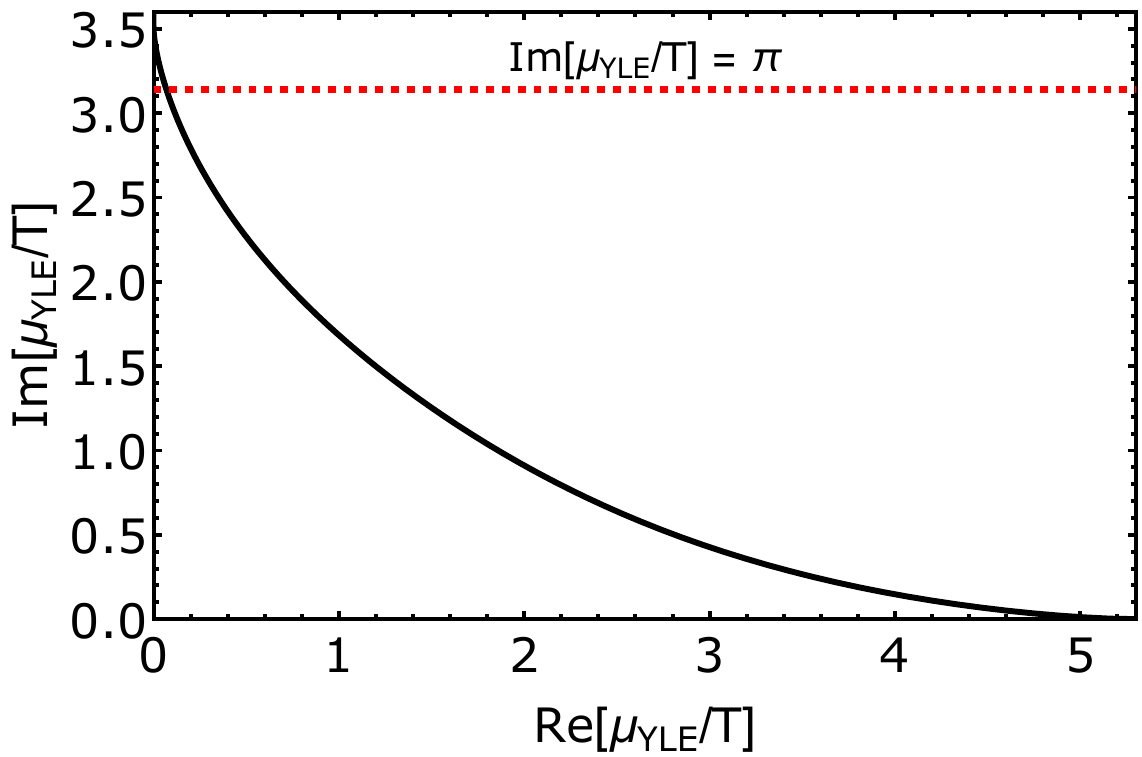}
    \caption{\justifying \textit{Left panel:} Lee-Yang Edge singularities plotted as a function of temperature in the imaginary plane. \textit{Right panel:} Lee-Yang Edge singularities on the real and imaginary $\mu_B$ axes displaying how the critical point in the imaginary plane is connected with the QCD critical point on the real axes.}
    \label{fig:YLE_singularities}
\end{figure*}

Lee and Yang showed that phase transitions of a thermodynamic system can be understood in terms of the complex singularities of the grand canonical partition function \cite{Yang:1952be,PhysRev.87.410}. The branch points of these singularities pinch the real axis at the point of second order phase transition, which is the critical point \cite{Basar:2023nkp}. Recently, there have been a lot of developments in obtaining these LYE singularities in the complex plane through lattice QCD simulations at imaginary $\mu_B$ \cite{Basar:2023nkp,Clarke:2024ugt,Adam:2025phc,Wan:2024xeu,Clarke:2026ifa} and in QCD-based models \cite{Savchuk:2019yxl,Basar:2021hdf,Zhang:2025jyv}. These Lee-Yang Edge singularities can just be treated as spinodals in the complex plane \cite{Skokov:2024fac} and hence can be obtained through the expansion of entropy density contours. 
We observe the spinodals using the conditions $dT/ds = 0 $ and Eq. \eqref{eq:TsO2}, but in this case $T_0$ and $\mu_B$ are in the imaginary plane, while the temperature $T$ is real. From this, we obtain the singularities at complex values of the chemical potential, the imaginary part of which is plotted as a function of temperature in the left panel of Fig. \ref{fig:YLE_singularities}.

The right panel in Fig.~\ref{fig:YLE_singularities} shows the Lee–Yang edge (LYE) singularities on the complex $\mu_B$ plane, illustrating how the critical point at imaginary chemical potential is connected to the QCD critical point at real $\mu_B$. A related analysis based on lattice QCD simulations was presented in Refs.~\cite{Clarke:2024ugt,Basar:2023nkp}, where a qualitatively different behavior of the singularity trajectories was observed. In particular, at high temperatures the LYE singularities were found to bend away from the Roberge–Weiss critical endpoint in the complex $\mu_B$ plane before approaching the line $\mathrm{Im}(\mu_B/T)=\pi$. In contrast, as shown in Fig.~\ref{fig:YLE_singularities}, our analysis suggests that the LYE singularities move from the real axis toward the Roberge–Weiss critical endpoint with increasing temperature, without an intermediate deviation in the complex plane. 

The red dashed line in the right panel of Fig.~\ref{fig:YLE_singularities} corresponds to $\mathrm{Im}\, \mu_B = \pi$, where the Roberge Weiss critical endpoint should be found with $\mathrm{Re}\,\mu_B /T = 0$. 
The $s$-contour expansion shows a critical endpoint at purely imaginary $\mu_B$, but it is located at $\mathrm{Im}\,\mu_B/T \approx 3.55$, due to the truncation of the expansion at second order in $\mu_B$ \cite{Shah:2024img}. Conversely, because of truncation errors, there is a small non-zero value of $\mathrm{Re}\,\mu_B/T$ for the singularity at $i\mu_B/T = \pi$. Regardless of truncation error, the $s$-contour expansion indicates that the CP in the real plane is connected to the Roberge-Weiss CP at imaginary $\mu_B$. Whether this behavior is a coincidence due to the truncated expansion or connected to actual physics is still under investigation, as there are no indications from lattice QCD or model studies that the two critical points should be connected to each other.

\section{Conclusions} \label{sec:conclusions}
In this work, we examined contours of constant entropy density in several QCD-inspired models to assess how the expansion developed in \cite{Shah:2024img} performs when truncated at $\mathcal{O}(\mu_B^2)$. Since the method relies on thermodynamic quantities becoming multi-valued in the region of a first-order phase transition, other quantities besides entropy density can also be used. Therefore, we further analyze the expansion by looking at contours of enthalpy density and energy density. Ideally, all of them should yield the same critical point, although truncation at second order can introduce differences. As a baseline test, the expansion is first applied to an ideal gas of massless quarks and gluons, which does not contain a critical endpoint. When the expansion was applied, it did not yield any CP and reproduced the contours of the model in both the imaginary and real $\mu_B$ planes.

However, when applied to the ideal hadron resonance gas model, which also does not have a CP, the expansion showed a fake critical endpoint at a high value of baryon chemical potential. The reason for this is that the actual contours of the model are not parabolic when $\mu_B$ approaches the mass of the proton, but rather become exponential. The expansion was also unable to capture the behavior of the contours at imaginary chemical potential, thus proving unreliable when applied to this particular model. The expansion was also studied in other versions of the  HRG model, including the excluded volume HRG model, the Van der Waals HRG model, as well as the cluster expansion model (which reduces to HRG at low temperatures).  
Because all these models behave very similarly to the ideal HRG at low temperatures and $\mu_B = 0$, a similar fake CP at an equivalently high value of $\mu_B$ with a $T_{0c} \leq 120$ MeV was predicted for all of them.
Given that the HRG model is expected to provide a good approximation of QCD thermodynamics at $T \lesssim 120$~MeV and $\mu_B = 0$, we conclude that the expansion of constant entropy contours is not reliable at $T_0 < 120$~MeV and any CP corresponding to $T_{0,c} < 120$ MeV is likely to be spurious.

Besides, the method produces a null result (no CP) when applied to the NJL model in its standard parameterization, which contains a CP at $(T,\mu_B) = (82,966)$ MeV. Our result here echoes the analysis in~\cite{Marczenko:2025znt}, where other parametrizations of the NJL model were considered.
Interestingly, the NJL model did not produce a fake CP at $T_{0,c} < 120$ MeV, unlike HRG-type models.
This likely reflects the absence of explicit hadronic degrees of freedom in the model.

Finally, the expansion was tested on models that predict a first-order phase transition in QCD and reproduce the available lattice QCD thermodynamics at $\mu_B = 0$. 
Firstly, it was applied to a holographic dual black-hole (BH) model, where the expansion successfully predicted a CP close to the actual CP location in the model. 
It also provided an accurate description of the true entropy contours of the model as the true contours in the model exhibited an almost quadratic behavior. 
The analysis of contours of constant energy or enthalpy density produces similar results, with the value of $\mu_{B,c}$ using energy density contours being closer to the actual value.
Another approach that describes lattice data at $\mu_B = 0$ and predicts a CP at finite density is the functional QCD (fRG-DSE) approach. 
Using tabulated data from Ref.~\cite{Lu:2025cls}, we constructed the expansion of entropy density contours at order $\mathcal{O}(\mu_B^2)$ and obtained a CP within 10\% of the actual CP value in this approach. 

Both the BH and fRG–DSE frameworks describe lattice QCD thermodynamics at $\mu_B=0$ and exhibit nearly quadratic constant-entropy contours, which justifies truncating the expansion at second order and helps explain why the corresponding critical-point estimates are close to the true values in these models. 
When applied to lattice QCD data~\cite{Shah:2024img}, the constant entropy method yields ($T_c,\mu_{Bc}) = (114,602)$ MeV, which is compatible with the CP location in BH and fRG-DSE frameworks.
Taken together, these results lend credence to the possibility that the QCD CP is located in this region of the QCD phase diagram.
Still, the possibility remains that this lattice-based estimate of the critical point in Ref.~\cite{Shah:2024img} does not correspond to a true critical point of QCD. 
Should this be the case, it would point to a limitation of the constant-entropy expansion when applied to lattice QCD, despite its successful application to the Holographic Blackhole and fRG–DSE approach.
Even though lattice QCD results indicate parabolic behavior of constant-entropy contours in the imaginary-$\mu_B$ plane (in the strangeness-neutral direction)~\cite{Borsanyi:2025dyp}, the breakdown of this behavior in real $\mu_B$ direction cannot be ruled out. 
Such a scenario would indicate an intrinsic difficulty of inferring critical behavior from effective descriptions constrained at small $\mu_B$ rather than an issue directly related to the shape of constant-entropy contours.

Combining the above observations within different models, one can give the following recipe to assess the reliability of CP predictions from the constant-entropy expansion method:
\begin{itemize}
    \item Inspect the behavior of the contours in the imaginary-$\mu_B$ plane (if available). If the contours show notable deviations from a quadratic behavior in $\mu_B$, the $\mathcal{O}(\mu_B^2)$ truncation is unlikely to be reliable.
    \item If the extracted value of $T_{0,c}$ is below $120$ MeV, the predicted CP is likely to be spurious and reflect the breakdown of the quadratic approximation in the hadronic regime.
    \item Cross-check the robustness of the inferred CP by comparing results from constant entropy, energy, and enthalpy density contours and verifying that they predict consistent crossings.
\end{itemize}

Finally, taking the quadratic behavior of constant-entropy contours from lattice QCD at face value, we reconstructed the full equation of state.
This was achieved by integrating the entropy density over the temperature at each value of $\mu_B$ and matching the integration constant to the HRG model at low temperatures.
As a result, we obtained a lattice-based equation of state model which contains a CP at $(T,\mu_B) \rightarrow (114,602)$ MeV in line with available constraints on the CP.
Alongside the equation of state, we studied the behavior of isentropic trajectories (contours of $s/n_B$) and their focusing in the vicinity of the CP.
We also studied the behavior of Lee-Yang edge singularities in the complex chemical potential plane.
For quadratic entropy density contours, the LYE trajectories smoothly connect the QCD critical point singularity on the real $\mu_B$ axis to the Roberge-Weiss endpoint at imaginary $\mu_B$.
This equation of state can be used to test the effect of a critical point in hydrodynamic simulations of heavy-ion collisions, and its predictions for baryon number fluctuations can be compared with measurements of proton number cumulants in heavy-ion collisions, albeit with an appropriate treatment of caveats associated with theory-to-experiment comparisons~\cite{Koch:2025cog}.
\\
\paragraph*{\bf Data Availability.} The data that support the findings of
this article are openly available \cite{shah2026data}
\vspace{1cm}
%
%
%
%
%
%
\section*{Acknowledgments}

We thank Yi Lu, Fei Gao and Jan Pawlowski for providing the data of fRG-DSE approach. M.H. was supported by the Brazilian National Council for Scientific and Technological Development (CNPq) under process No. 313638/2025-0.
V.V. was supported by the U.S. Department of Energy, 
Office of Science, Office of Nuclear Physics, Early Career Research Program under Award Number DE-SC0026065.
This material is based upon work supported by the National Science Foundation under grants No. PHY-2208724,
PHY-2116686 and PHY-2514763, and within the framework of the MUSES collaboration, under Grant No. OAC-2103680. This material is also based upon work supported by the U.S. Department of Energy, Office of Science, Office of Nuclear Physics, under Award Numbers DE-SC0022023, and DE- SC0023861, as well as by
the National Aeronautics and Space Agency (NASA) under Award Number 80NSSC24K0767.

\bibliography{main}
\end{document}